\documentclass[final,1p,times]{elsarticle}

\usepackage{graphicx}
\usepackage{amssymb}

\usepackage[utf8]{inputenc}
\usepackage{float}
\usepackage{floatpag}
\usepackage{soul}
\usepackage{caption}
\usepackage{tabularx}
\usepackage{paralist}
\usepackage{listings}
\usepackage{xcolor}
\usepackage{enumerate}
\usepackage{multirow}
\usepackage{dblfloatfix}
\usepackage[intlimits,sumlimits,namelimits,centertags]{amsmath}
\usepackage{url}
\usepackage{eurosym}

\newcommand{\boxedtext}[1]{\fbox{\scriptsize\bfseries\textsf{#1}}}
\newcommand{\nota}[2]{
	\boxedtext{\small #1}
	{\small$\blacktriangleright$\emph{\textsl{#2}}$\blacktriangleleft$}
}
\newcommand{\todo}[1]{\nota{\hl{TODO}}{\hl{#1}}} 
\renewcommand\todo[1]{} 

\newcommand{\sbaltes}[1]{\textcolor{blue}{#1}}

\renewcommand{\sbaltes}[1]{}

\newfloat{atticElem}{tbp}{loaE}[section]

\newcommand{\attic}[1]{\begin{atticElem}[p]\noindent\fbox{
			\begin{minipage}{\linewidth}
				\small {#1} \end{minipage}}\end{atticElem}}
\renewcommand{\attic}[1]{} 

\makeatletter
\newcounter{patterncount}
\lstnewenvironment{pattern}[2][]
{

	\lstset{language=Java,firstnumber=auto,name=patterncount,
		mathescape=true,escapeinside={(*}{*)},basicstyle=\ttfamily,#1}}
{\addtocounter{patterncount}{1}}
\makeatother

\journal{Journal of Systems and Software}

\begin{document}

\begin{frontmatter}


\title{On the Diversity and Frequency of Code Related to Mathematical Formulas in Real-World Java Projects}



\author[L1]{Oliver Moseler*}

\author[L1]{Felix Lemmer}

\author[L2]{Sebastian Baltes}

\author[L1]{Stephan Diehl}



\address[L1]{Computer Science, University of Trier, Germany}
\address[L2]{School of Computer Science, The University of Adelaide, Australia}


\begin{abstract}
In this paper, the term formula code refers to fragments of source code that implement a mathematical formula. We present empirical studies that analyze the diversity and frequency of formula code in open-source-software projects. In an exploratory study, we investigated what kinds of formulas are implemented in real-world Java projects and derived syntactical patterns and constraints. We refined these patterns for sum and product formulas to automatically detect formula code in software archives and to reconstruct the implemented formula in mathematical notation. In a quantitative study of a large sample of engineered Java projects on GitHub we analyzed the frequency of formula code and estimated that one of 700 lines of code in this sample implements a sum or product formula. For a sample of scientific-computing projects, we found that one of 100 lines of code implements a sum or product formula. To assess the need for tool support, we investigated the helpfulness of comments for program understanding in a sample of formula-code fragments and performed an online survey. Our findings provide first insights into the characteristics of formula code, that can motivate further studies on the role of formula code in software projects and the design of formula-related tools.

\attic{In this paper, the term \textit{formula code} refers to fragments of source code that implement a mathematical formula. We present empirical studies that analyze the diversity and frequency of formula code in open-source software projects. First, in an exploratory study, we investigated what kinds of formulas are implemented in real-world Java projects and derived syntactical patterns and constraints for reoccurring formula code fragments.
 We refined the patterns for sum and product formulas to not only automatically detect formula code in software archives, but also to reconstruct the implemented formula in mathematical notation. Second, in a quantitative study of a large sample of engineered Java projects on GitHub we analyzed the frequency of formula code and estimated that one of 700 lines of code in this sample implements a sum or product formula. We repeated the study with a sample consisting solely of scientific-computing projects, found a formula code density that was 7.4 times higher and estimated that one of 100 lines of code implements a sum or product formula. Furthermore, to assess the need for tool support related to formula code, we investigated the helpfulness of comments for program understanding in a small sample of formula code fragments and performed an online survey where participants had to detect defects in formula code.
  Our findings provide first insights into the characteristics of formula code, that can motivate further studies on the role of formula code in software projects and the design of formula-related tools.}

\attic{
We introduce the term \textit{formula code} to refer to fragments of source code that implement a mathematical formula. Since there exists a wide range of mathematical formulas and their implementations, in this paper, we mainly focus on code fragments computing  numerical values in a way that can be expressed by sum and product formulas in common mathematical notation. We present empirical studies that analyze the diversity and frequency of formula code in open-source software projects. First, in an exploratory study, we investigated what kinds of formulas are implemented in real-world Java projects and derived syntactical patterns and constraints for reoccurring formula code fragments. We refined the patterns for sum and product formulas to be able to automatically detect implementations of these formulas in software archives. Our approach is not only able to classify source code as formula code for sums and products, but also to reconstruct an underlying formula in mathematical notation. We performed a quantitative study analyzing the frequency of formula code on GitHub. Overall, we estimate that one of 700 lines of code in our sample of engineered Java projects 
is part of the implementation of a sum or product formula. Furthermore, based on our analysis we roughly estimate that every 13th simple (non nested) and every 25th nested \texttt{for}-loop implements a sum or product formula. Our investigation of the application domains of projects having a high formula code density reveals that there is a major intersection with the domain of scientific-computing. Thus, we repeated the study with a sample consisting solely of scientific-computing projects and found a formula code density that was 7.4 times higher compared to the one of the previous sample. Thus, we estimate 
that one of 100 lines of code in our sample of scientific computing Java projects is part of formula code and that every 4th simple, and every 16th nested \texttt{for}-loop implements a sum or product formula. While the implementations of formulas in Java are quite different in the control and data structures used, we found that sum and product formulas, in particular the ones implemented through simple \texttt{for}-loops, make up a large part of those implementations.  Our findings provide first insights into the characteristics of formula code, that can motivate further studies on the role of formula code in software projects and the design of formula-related tools, e.g. debugging and code comprehension tools, as well as language features. 
}
\end{abstract}


\end{frontmatter}

\newpage
\begin{center}
{\Large 
On the Diversity and Frequency of Code Related to Mathematical Formulas in Real-World Java Projects
}
\end{center}



\section{Introduction}
Since there exists a wide range of mathematical formulas and their implementations, in the context of this paper, we use the term \textit{formula code} to denote fragments of source code that compute a numerical value (scalar, vector, matrix) in a way that can be expressed in a common mathematical notation~\cite{Cajori:1929} or by a mixed form of source code artifacts and maths symbols.

The correct and performant implementation of mathematical aspects is a crucial part influencing the success of a software system.
The destruction of the Mariner 1 spacecraft in 1962 caused a \$18.5 million financial damage due to a faulty implementation of a mathematical formula in FORTRAN, in particular a missing superscript bar `signifying a smoothing function, so the formula should have calculated the smoothed value of the time derivative of a radius'~\cite{ComputerWorldCom, PlanetaryProbeHistory, TheRisksDigest}. This is only one prominent example where formula code happens to be a critical part of a software system. 
An incident during Qantas Flight 72 in 2008 shows that it is often not the formula implementation alone that may cause potentially catastrophic outcomes, but the assumptions being made when specifying the intended behaviour.
On that flight, the way in which the Airbus A330-300’s flight control computer determined an important flight parameter based on the average and median values of three sensor outputs plus a combination of heuristics to deal with spike values 
caused the airplane to enter a steep descent, injuring passengers and crew members~\cite{atsb:2008}.

 The general discipline of software engineering is studded with problem solving and therefore with mathematical reasoning~\cite{DBLP:journals/cacm/Henderson03}. Thus maths is omnipresent in the work of a software engineer and represented in the software systems' code base by the utilization of one or more programming languages.
 Furthermore, the early detection of defects in maths implementations and discovery of opportunities to increase the software's performance within the development of a software system could save hours of testing and consumption of resources, especially in long running computational environments, such as scientific or high performance computing. Software development tools supporting the implementation and comprehension of formula code would not only be of advantage in such specialized scientific domains.
To some extent every computer program contains at least some logic or discrete mathematics like combinatorics, probability theory, graph theory or number theory.

In psychological studies Landy et al. investigated the importance
of spatial relationships in mathematical notations~\cite{Landy:2014}.
For example, test persons falsely rated the equation $a + b * c + d = c + d * a + b$ 
valid, if the distance between symbols did not correspond to the
operator precedence. The authors concluded `that competent symbolic reasoners typically rely on semantically irrelevant properties of notational formulae in order to quickly and accurately—but also sometimes inaccurately—solve symbolic reasoning problems.' 
Thus, we assume that a mathematical representation is beneficial in order to reduce both the time for comprehending the code as well as assessing its correctness. In mathematical notation, operators and symbols are arranged in two dimensions allowing a more compact view.  


When we started this project, we found ourselves in the situation that we could not find much work to build on.
Surprisingly, although mathematical formulas obviously play an important role in programming, so far, software engineering research has not empirically studied the implementation of formulas in common programming languages nor developed much tool support for implementing, debugging and optimizing formulas. Research as well as tools have rather focused on different levels of program abstractions (e.g. statement, class, or package level) or on higher-level abstractions (e.g. features, aspects, or components). \todo{SD: Referenzen?}
In general purpose programming languages such as Java, Python or C++, formula code does not simply correspond to a syntactical category such as expression: Not all expressions in a program implement a formula (e.g. \verb+fopen(fname)!=-1)+ and not all formulas are implemented as expressions (e.g. program code for $ \sum_{i=1}^n a_i $ typically uses {\tt for}-loops). 
Identifying formula code and making software developers aware of recurring patterns, best practices, pitfalls, and useful 'hacks' related to the implementation of mathematical aspects as well as simply showing the code in maths like notation can help to reduce development time and technical risk, as well as the effort for code comprehension and communication, particularly in cross disciplinary development teams, where experts of mathematically predominant domains, e.g. mathematicians or chemical scientists, and software developers tightly work together. 

Our vision is that better understanding of the characteristics of formula code will help to develop novel tools, beyond formula editors, for understanding, maintaining, debugging and optimizing formula code as well as to design new APIs and language features to enhance a software systems maintainability and thus its overall code quality.
\todo{SD: Tools fuer Synthese, Editoren existieren !!!}

\attic{Neither did we find relevant work that would discuss any notion of formula code, nor did we find relevant work on how to automatically detect it.}

\attic{
But there exists an analogy to the research work on code clones. While code clones are generally defined as code fragments which are `similar by some given definition of similarity', tools and studies on code clones considerably differ in the similarity measure they use to define and detect code clones~\cite{Roy:2009}. In our case, we define formula code as code fragments implementing a computation of a numerical value which can alternatively be described through a, in mathematics commonly used, notation. Thus, our definition depends on what we consider to be a computation of a numerical value and when we consider a code fragment to implement a computation of a numerical value\attic{what we consider to be a code fragment implementing such a computation of a numerical value}. 
}

To gain insights on the use of formula code in real software projects and as 
a basis for future research on tool support and language design, we want to answer the following two research questions:

\begin{description}
\item [RQ1 (Diversity):] \textit{What kinds of formula code occur in
   real-world software projects?}
\item [RQ2 (Frequency):] \textit{How frequent is formula code both at file and line granularity?}  

\end{description}

\attic{nicht in diesem Papier RQ4: Differences between formula code
      and the remaining code.
      }

To answer these research questions, we performed two studies, one qualitative and one quantitative study. For our qualitative study, we ran a keyword-based search to find formula code in open source Java projects on GitHub (Section~\ref{keyword:sec}). While this keyword-based approach suffers both from low recall and low precision and requires a lot of manual post-processing, it helped us to find an initial set of real-world formula code samples, that we then manually analyzed in order to gain first insights on the kinds of formula code that exist (RQ1) and to derive patterns of formula code (Section~\ref{patterns:sec}).

\sbaltes{Schaut euch folgenden Absatz mal an (bzgl. "the focus on the Java programming language that according to R2 is not predestined for formula code"}
While there exist programming languages such as FORTRAN or MATLAB that are tailored for numerical computation, even in those languages there exists a conceptual gap between imperative language features and the declarative nature of mathematical formulas.
Nevertheless, imperative languages such as Ada are still commonly used to implement mathematical formulas, for example in the aviation industry~\cite{ada:boeing,ada:airbus}. 
However, mathematical formulas do not only play a role in safety-critical domains like aviation, they are also central to many concepts in computer graphics, machine learning or finance. \sbaltes{Haben wir eine Referenz fuer ersteres Statement und vielleicht ein zweites gutes Beispiel?}
We decided to focus on a general-purpose programming language, Java, because not having direct support for implementing mathematical formulas, for example by having a matrix datatype and related operations, adds an additional indirection layer that makes it even harder for developers who need to implement mathematical formulas in such languages.
Our study has shown that, while relatively rare, mathematical formulas are being implemented in Java. In some of our sampled projects more than 5\% of source code implemented a sum or product formula and every 4th loop in the scientific computing projects 
implemented such a formula. \sbaltes{Können wir hier auf irgendein interessantes Beispiel/Ergebnis vorgreifen?}

Given the diversity of the formulas that we found, we decided to focus on sum and product formulas for our quantitative analysis, because they are structurally non-trivial and we expected them to occur quite frequently. We will use the term \emph{SP-formulas} in the rest of this paper to refer to sum and product formulas and the term \emph{SP-formula code} to refer to source code we can express as SP-formulas in a mathematical notation\attic{that implements SP-formulas}.
The derived patterns from the quantitative study form the basis of our pattern-based method for detecting SP-formula code (Section~\ref{tool:sec}). Our approach is not only able to classify source code as SP-formula code, but also to reconstruct the formula 
implemented by the code in mathematical notation\attic{the underlying formula}. 
Our evaluation shows that the pattern-based method has a very high precision (almost 100\%) and a modest recall (31\%) for SP-formula code. Moreover, for 85\% of the detected SP-formula code the reconstructed formula was both correct and completely described the computation of the matched code.
Finally, to answer research question RQ2, we applied the tool on a sample of 1000 different open source Java projects to detect SP-formula code on GitHub (Section~\ref{casestudy:sec}).
For a smaller sample, we also compared the densities from arbitrary application domains with those in scientific computing.

Furthermore, to assess the need for tool support related to formula code, we performed two small studies. First, we investigated the source-code comments of 139 real-world formula code fragments and observed that they are rarely commented in a way that helped to better comprehend the implemented formula. Second, in an online survey, we examined whether showing a reconstructed mathematical formula next to the formula code supports the defect detection task. For the survey we selected four real-world formula code fragments of different complexity. 
Due to the low number of participants that actually completed the survey, 
the quantitative results comparing their performance with and without showing
the reconstructed formula are not conclusive.
However, for complex formulas the accuracy scores considerably decreased, which offers ample room for tool development. This adds to our assumption that tools can improve development tasks related to formula code. 
Moreover, the vast majority of the participants found a tool that reconstructs the mathematical formula from the source code as helpful.

In summary our contributions are: A first qualitative study investigating the nature 
of formula code; a set of recurring formula code patterns derived from the findings in the qualitative study; a quantitative study on the density of sum and product formula code in open source Java projects.

\todo{SD: besserer Begriff fuer modest}

\todo{SD: Explizit die Contributions auflisten. NEIN, KEIN PLATZ}

\todo{OM: Motivation: Famous Formula Code Bugs like Mariner 1 to emphasize the importance of formula code}

\todo{OM: Motivation: Visual mathematical representation to support cross disciplinary communication, i.e. program comprehension}

\todo{OM: Motivation: Formula code patterns. Analogous to Design Patterns. When people are aware of those patterns it will increase program comprehension. Ref to a paper where people showed that with design patterns}

\attic{
\todo{SD: der folgende Absatz sollte später verwendet werden, wenn die Beschränkung
 auf Java diskutiert wird.}
The programming language FORTRAN  (FORmula TRANslation) is trimmed on implementing numerical computations and thus is one example of a programming language supporting necessary abstractions for implementing mathematical formulas. There are other specialized languages like MATLAB or Mathematica which come with data types for matrices and vectors and therefore try to lower a developers effort to implement a formula also in the way that syntactical elements of the language lean towards the well known mathematical notation of the respective formulas. But there are obvious limitations since program code is written line wise and each character written has to fit in this line vertically as well as horizontally. Hence even elements of a very simple formula in mathematical notation, e.g. $\sqrt[]{x}$, cannot be coded right away but with a respective function call or even much more code. We note that this will increase the number of symbols used to implement a formula in a programming language compared to its representation in mathematical notation. \todo{OM: psycho reference SD: Die beziehen sich auf graphische Darstellung, die ist nicht Teil dieser Studie} Based on that we assume that it takes more mental effort to understand formulas implemented in a programming language than in mathematical notation and thus it is possible to enhance the program comprehension process. \todo{OM: More on program comprehension, developers spend more time in reading than writing code?} In the long term we will address this by creating tools supporting the creation of a more mathematical notation like representation of formula code up to and including debuggers specialized in formula code debugging requirements which have to be acquired first. 
}

\attic{
\todo{what is formula code?}
\todo{OM: the following is wrong, but!}
The visual representation of formula code, either in mathematical notation or in a programming languages syntax, leads to a guideline on the decision on what we already refer to as formula code and what we can denote as clearly no formula code although only on the former. Since there doesn't exist a straightforward definition of a mathematical formula, almost everything can be formula code in the definition mentioned earlier. So we need to clarify the distinction between formula code which we want to address in our work and which not. \todo{OM: move to qualitative study?} Therefore we decided to denote program code as to be formula code if we are able to create an appropriate mathematical representation and it significantly differs from its program code representation. For example, {\tt a+2} obviously is an implementation of a trivial mathematical formula but we do not search for program code like this. On the other hand the program code {\tt a/3}  clearly differs from its mathematical representation since it 
would look like {$\frac{a}{3}$}.
}

\section{Related Work}
\label{relwork:sec}
Work on automatic layout of formulas based on textual specifications as well as graphical formula editors~\cite{Kernighan:1975,Knuth:1979,Levison:1983} dates at least back to the seventies. There is also a considerable amount of research 
on producing internal representations from graphical ones using image
recognition techniques~\cite{Zanibbi:2012, Chan:2000}. 

\attic{
The framework MPS (\textit{Meta Programming System}) by Jetbrains supports the design of domain specific languages. This includes corresponding individual editors for those languages, which are capable of overcoming the traditional one dimensional way of coding to enable domain experts to write programs in their familiar notation. As an example, they designed a formula language with an editor where a programmer forms a program by using a mathematical like notation in a two dimensional layout ~\cite{JetbrainsMPS2018}. }  

In the area of programming, recent work includes the integration of a formula editor in a common IDE as a domain specific language extension using Jetbrains MPS~\cite{JetbrainsMPS2018}.

Since the advent of mining software repositories, researchers have developed numerous methods for analyzing software repositories to detect patterns in the source code (e.g., aspects\cite{DBLP:conf/msr/BreuZL06} or API patterns~\cite{DBLP:conf/msr/XieP06}) or its changes (e.g., recommended changes~\cite{Zimmermann-et-al:tse} or refactorings~\cite{WeissgerberDiehl:06}).

The only work on reconstructing mathematical formulas from source code has been published by Moser et al.~\cite{Moser2016,Moser2015}. Their RbG tool extracts formulas from annotated source code to produce a documentation of the source code. The tool was developed for Fortran and C++, requires manual annotations, uses static program analysis and covers only a small part of possible formulas. 

While there exist tools for searching mathematical formulas in documents (e.g., using tree-edit distance \cite{Kamali:2013}) or for searching mathematical expressions in software binaries~\cite{DBLP:conf/msr/JainPVP08} (using fingerprints) in order to build a search system which is capable of querying for software libraries implementing a mathematical term, to our knowledge, so far, no methods for detecting formula code in software repositories or empirical studies on formula code in common programming languages have been published.
To some extent, almost every implementation of a mathematical formula is working with numerical values such as integer or floating point numbers.

The computation of numerical values comes with its own challenges and pitfalls. Recently, studies on characteristics such as categories, symptoms, frequencies and possible fixes of numerical bugs have been conducted by Di Franco et al.~\cite{DBLP:conf/kbse/FrancoGR17}. 
They have neither investigated patterns in the code of numerical computations nor tried to detect it automatically, 
but focused on bugs which are related to these code fragments.
While parts of numerical computations can be described by
formulas, they are often more algorithmic in nature.

\todo{SD: In einer Studie
aus dem Jahr 2017 haben Di France et al. Charakteristika numerischer Programmierfehler (bugs)  in fuenf haeufig genutzten, numerischen Softwarebibliotheken untersucht. Sie klassifizieren die Fehler in Accuracy Bugs,
Bugs Related to Special Values (z.B. NaN, infinity), Convergence Bugs und Correctness Bugs, wobei die letzte Klasse mit 37\% am haeufigsten vorkam.}

\section{Keyword-Based Search for Formula Code}
\label{keyword:sec}

To get first insights into what kinds of formulas are actually implemented in real-world software projects (RQ1) we conducted a qualitative study using GitHub as our data source. With more than 96 million repositories and 31 million contributors involved (as of September 2018 \cite{GitHubOctoverse2018}), GitHub is one of the most popular code hosting platforms today. It is not only used by developers for their personal projects, but also by large companies such as Google, Microsoft, or Facebook.
We restricted our search to projects containing Java source code, since Java is one of the most popular general purpose programming languages according to the IEEE language ranking (as of July 2018~\cite{IEEELangRank2018}) and the TIOBE index (as of October 2018~\cite{TIOBEIndex2018}). Our decision to investigate Java projects is based on the assumption that they contain code that was developed by programmers of all levels from novices to experts 
with respect to their programming as well as maths skills.

\todo{SD: Wuerde den Satz streichen: Although it is widely used in maths rich domains like scientific computing, we decided against investigating projects implemented with the Python programming language because missing type information, vectorization and broadcasting can easily lead to misinterpretation of the code semantics. 
}

\attic{
Our choice of Java is based on the assumption, that it is one of the most popular general purpose programming languages and thus experts in the field of implementing mathematical formulas, e.g. physicist or chemists, may not make use of it. If we would consider MATLAB or a comparable specialized mathematical programming language, we surely would discover many implementations of mathematical formulas but only of people who are also specialists in the field. We want to track down code fragments of software developers who may or may not be familiar with the formula they have to implement. To be able to create tools which help software developers implementing mathematical formulas, we first contemplate the non experts. We are of the opinion that for the moment we can retrieve more requirements to potential developments tools this way than to consider the experts in the field first. Another reason to not start with mathematical tinged programming languages is the open source availability. Most software projects in the domain of scientific computing are closed source albeit there are a few available like \todo{OM: find at least one}.
}

To find formula code, we searched for certain keywords in the commit messages and code comments, because software developers use these annotations to document and communicate various aspects of source code artifacts and changes, including their intent.
For our study, we wanted to find program code that a developer documented to be associated with mathematical formulas. Hence, we searched for occurrences of the keywords \textit{`formula'}, \textit{`equation'}, \textit{`math'}, \textit{`theorem'}, \textit{`sum of'} and \textit{`product over/of'} within the Git commit logs and comment sections of Java code \attic{ hosted on GitHub}. While our list of keywords is certainly not comprehensive, it allowed us to find a sufficiently large and divers set of samples for our subsequent manual analysis.

In a first attempt, we utilized the Boa language and infrastructure~\cite{Dyer2013} to retrieve candidates. We conducted a search on the `2015 September/GitHub' dataset available through Boa. This approach revealed only a modest number of useful hits. Furthermore it was a time consuming task to deduce the interesting code fragments, if existent, from the commit log messages since a commit usually relates to multiple files and thus required a significant amount of manual effort to identify the relevant file and source-code fragment in this file. Thus we changed our strategy and
only searched in the source code comments, because the comments are in the same file and usually very close to the source code fragments that they annotate. Unfortunately, Boa did not provide access to source code comments, since they were not contained in its data model.
As a consequence, we switched to  Google Big Query~\cite{GoogleBigQuery18} (GBQ), a web service 
for searching GitHub using the GBQ GitHub and GHTorrent datasets, 
that allows executing SQL-queries to search for keywords within source code comments of Java files.

\attic{
Thus, we changed our strategy and decided to search in the source code comments, because the comments are in the same file and usually very close to the source code fragments that they annotate. 
}

\attic{
Since it is a time consuming task to deduce the interesting code fragments, if existent, from the commit log, we focused on the strategy to search in the source code comments, because the comments are in the same file and usually very close to the source code fragments that they annotate.  
}
\attic{
Unfortunately, Boa did not provide access to source code comments, since they were not contained in its data model. Consequently, we switched to Google Big Query ~\cite{GoogleBigQuery18}, a web service  which allows to search GitHub (i.e. actually the GH-Torrent) with SQL-queries and even search for keywords within source code comments of Java files. 
}


\todo{OM: Mention the used regex in GBQ and that it searches line wise so there are limitations in finding comment sections!}

\subsection{Exploratory study}
With the help of GBQ, we created one big CSV file containing all matches of each keyword mentioned above. Each match is recorded as a tuple with the following data:
\begin{center}
\texttt{(id, match, link, line, repo\_name, path)}
\end{center}
where \texttt{id} denotes a file identifier in the GBQ GitHub dataset, \texttt{match} the matched keyword, \texttt{link} the link to the respective repository on GitHub, \texttt{repo\_name} the corresponding repository name consisting of the account as well as project name and \texttt{path} denotes the path of the file containing the match.




The goal of our qualitative analysis was to find common properties of real-world formula code. To this end, we followed an iterative analysis process borrowing some ideas (open coding, iteration, theoretical sampling and saturation) from Grounded Theory ~\cite{GlaserStrauss:67}:
\begin{enumerate}
\item Compute a set of matches (sample) using feedback from the
     previous iteration to adapt the sampling strategy and collect data 
     that will more likely lead to new insights (theoretical sampling).
\item For each match:
\begin{itemize}
\item  Decide whether the source code fragment implements a formula
\item  Try to reconstruct the underlying mathematical formula, 
       observe and describe phenomena of the formula and the formula code (open coding).
\end{itemize}
\item If the analysis of the sample lead to new insights (i.e. new or refined codes) then
      repeat the analysis with another sample (Step 1 ). Otherwise, 
      our codes cover all relevant phenomena (theoretical saturation).
\end{enumerate}

\todo{R2: research methodology, iterations, keywords. Zwischen den Samples einbauen.}

\attic{ The comments themselves sometimes gave hints to a concrete formula in the literature. We looked those up and compared them to our interpretation. Most of the time these were quite similar or mathematically equal to the formula we derived from the code \todo{OM: example?}. In the case we had not been able to reconstruct a formula but the hints the software developers made sure that the program code is implementing a mathematical formula, we also included the example in our coding. }
Following the above process, we conducted 4 iterations of group sessions \attic{including the authors}until we reached a saturation on our observations.
In the first iteration, we used the keywords \textit{`theorem'}, \textit{`formula'}, 
\textit{`equation'}, and \textit{`sum of'} and ranked the projects by the number of matches and started to manually inspect the matches of the top ten projects and tried to reconstruct the underlying formula.
Very quickly it became apparent that the majority of these matches \attic{ (actually 60,815 matches)} were commented out Java code fragments with references to the class \texttt{java.lang.Math}, in particular 
calls to functions of this standard library like \texttt{Math.sqrt(...)}
(see supplementary material ~\cite{moseler_oliver_2018_1252324}). So, we actually had many false positives and we stopped coding after looking at 12 matches. In these matches we observed already 16 of the 17 properties that are shown in Table~\ref{coded:tab}. 
To reduce the number of false positives, we removed the keyword \textit{`math'} from the query and computed for the second iteration a sample which we sorted by popularity to filter small toy projects. 
\todo{SD: verbreitete Praxis, Referenz?}
We coded another 21 matches and observed only one additional property, namely co-index variables. For the third iteration we decided to search for the keywords \textit{`product of'} and \textit{`product over'}, because we thought that they are computed in a similar way as sums, but we wanted to check whether there are any differences except for the multiplication operation in their implementations. In contrast to the keyword \textit{`sum of'}, for the keyword \textit{`product of'} we got a lot of false positives, thus we coded only 4 including one \texttt{dot product of} of two vectors: {\tt float dot = (crossX * dx + crossY * dy); } We did not observe any new properties. To make sure, that we did not miss relevant phenomena because of our choice of projects and ranking of the matches, we searched for all keywords in a random selection of projects and randomly chose 10 matches that we coded. Again, we did not observe any new relevant properties.

In summary, we used the following four samples for our qualitative analysis:
\begin{description}
\item [\sf Top10M:] Matches of top 10 projects sorted by number of matches per project (keywords:
  theorem, formula, math, equation, sum of)
\item [\sf Top50P1:] Matches of top 50 projects sorted by popularity (keywords: theorem, formula, equation, sum of)
\item [\sf Top20P2:] Matches of top 20 projects sorted by popularity (keywords: product of, product over)
\item [\sf Random:] Matches randomly selected (keywords: theorem, formula, equation, sum of, product of, product over)
\end{description}



\attic{ The samples {\sf TOP50P1} and {\sf TOP20P2} were created by sorting the projects by number of watchers as we think of this as a reasonable indirect measure of popularity to reduce noise in our sample and thus filter out toy projects. We took the top 50 and top 20, respectively to define each sample. }

\begin{figure}
	\centering
	\includegraphics[width=0.8\textwidth]{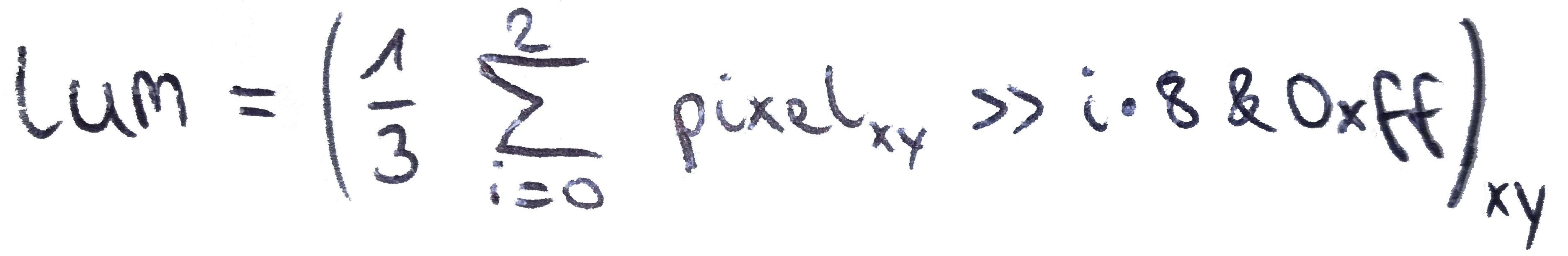}
   	\caption{\label{lumformula:fig}Formula reconstructed from code in Listing~\ref{lumlisting:lst}}
\end{figure}

In total, we closely inspected 142 matches from 101 different open source Java projects on GitHub and found 47 matches from 21 different projects to be formula code. For those matches, we reconstructed the
underlying mathematical formula of each fragment (example shown in Figure~\ref{lumformula:fig}). The observed and coded phenomena include control flow statements, roles of variables, 
comprehensibility and mathematical data type. Note, albeit we consider the Java language feature of lambda expressions being close to a mathematical notation, it did not occur once in our qualitative study. For reconstructing the formula, the comments were rarely helpful. In most cases the comments simply stated what was computed, e.g. {\tt // Calculate chi-square scores}. In 5 cases the comments contained a reference to the literature or the web. In one case, the formula was explicitly given in the comment. In two cases only meta-information was given in the comment ({\tt // It’s a weird formula, but tests prove it works.}, {\tt // formula below does not work with very large doubles}). 





\attic{
\begin{table}
\caption{Comprehensibility properties in coded samples~\label{comprehensibility:tab}}
\begin{tabular}{lrrrrr}  
{\bf Sample} & {coded} & context & expressible & proximity & mix-form \\
\hline
Top ten (\#matches) & 12 & 7  & 11 & 9 & 3\\
Top 50 (popularity) & 25 & 14 & 22 & 15 & 9 \\
Random              & 10 & 7  & 10 & 7 & 6\\
\hline
Total               & 47 & 28 & 43 & 33 & 18
\end{tabular}
\end{table}
}

\attic{
\begin{table}
\caption{Properties of the coded samples~\label{coded:tab}}
\begin{tabular}{llrrrrr}
\hiderowcolors
 & Sample & \rotatebox{90}{\sf Top10M} & 
            \rotatebox{90}{\sf Top50P1} & 
            \rotatebox{90}{\sf Top20P2} & 
            \rotatebox{90}{\sf Random}  & Total\\
\hline
& {coded}         & 12 & 21 & 4 & 10 & 47 \\
\hline
\multirow{4}{*}{\rotatebox{90}{\parbox{1.5cm}{\centering \small control-flow\\ statements}}} 
&{\tt \bf for}   & 4  & 10 & 3 & 7  & 24 \\
&{\tt \bf while} & 1  & 1  & 0 & 0  & 2 \\
&{\tt \bf if}    & 5  & 8  & 0 & 3  & 16 \\
&incremental     & 1  & 2  & 0 & 0  & 3 \\
\hline
\multirow{4}{*}{\rotatebox{90}{\parbox{1.5cm}{\centering \small roles of\\ variables}}}
&read only       & 9  & 17 & 4 & 9  & 39 \\
&accu            & 3  & 12 & 3 & 4  & 22 \\
&index           & 5  & 13 & 3 & 7  & 28 \\
&co-index        & 0  & 2  & 0 & 1  & 3 \\  
\hline
\multirow{4}{*}{\rotatebox{90}{\parbox{1.5cm}{\centering \small comprehen-\\sibility}}}
&w/o context     & 7  & 11 & 3 & 7  & 28 \\
&expressible     & 11 & 18 & 4 & 10 & 43  \\
&proximity       & 9  & 11 & 4 & 7  & 31 \\
&mix-form        & 3  & 8  & 1 & 6  & 18 \\  
\hline
\multirow{4}{*}{\rotatebox{90}{\parbox{1.5cm}{\centering \small mathemati-\\cal data type}}}
&matrix       & 4  & 5 & 1 & 2  & 12 \\
&array        & 9  & 7 & 3 & 4  & 23 \\
&series       & 2  & 8 & 3 & 2  & 15 \\
&point/vector & 5  & 8 & 3 & 6  & 22 \\  
\hline
\end{tabular}
\end{table}
}





\paragraph{Control-Flow Statements}
As shown in Table~\ref{coded:tab}, 24 of the 47 verified formula code fragments made use of a \texttt{for}-loop of which 12 were simple non nested \texttt{for}-loops, eight were double nested \texttt{for}-loops (one nested into another) and four were triple nested \texttt{for}-loops. We only found two examples where \texttt{while}-loops were used to implement a mathematical formula---one was non nested and the other double nested. 
Furthermore, our findings revealed that most sum and product formulas were implemented using \texttt{for}-loops, which is not surprising as \texttt{for}-loops are closer to the mathematical representation than other loops. The same holds for formulas which required iterating
 through a vector or matrix.
The conditional statement (\texttt{if}) was found 16 times, either within a loop body to function as a filter (seven times) or completely outside of any loop (nine times). 
In three cases, we also determined  an incremental computation which means that, at runtime, a series of method calls resulted in a more complex implementation of a mathematical formula---the formula actually describes the invariant of the value of a variable. For example, the method {\tt put(float value)} in the class {\tt FloatCounter} of the project {\tt libgdx/libgdx} (Listing~\ref{floatcounterlisting:lst}) incrementally updates simple statistical values, such that {\tt average} equals
$\frac{1}{n}\sum_{i=1}^{n} v_i$ where $n$ is the current value of {\tt count} and $v_i$ are the actual values of the parameter {\tt value} of all invocations of the method: 

\begin{table}
	\caption{Properties of the coded samples~\label{coded:tab}}
	\centering \small
	\begin{tabular}{lr|cccc|cccc|cccc|cccc}
		\ & \ & \multicolumn{4}{c|}{control-flow}
		& \multicolumn{4}{c|}{}
		& \multicolumn{4}{c|}{}
		& \multicolumn{4}{c}{mathematical} \\
			\ & \ & \multicolumn{4}{c|}{statements}
		& \multicolumn{4}{c|}{roles of variables}
		& \multicolumn{4}{c|}{comprehensibility}
		& \multicolumn{4}{c}{data type} \\
		\cline{3-18}
		Sample &  coded
		&\rotatebox{90}{{ \bf for}}
		&\rotatebox{90}{{ \bf while}}
		&\rotatebox{90}{{ \bf if}}
		&\rotatebox{90}{ incremental}
		&\rotatebox{90}{ read only}
		&\rotatebox{90}{ accu}
		&\rotatebox{90}{ index}
		&\rotatebox{90}{ co-index}
		&\rotatebox{90}{ w/o context}
		&\rotatebox{90}{ expressible}
		&\rotatebox{90}{ proximity}
		&\rotatebox{90}{ mix-form}
		&\rotatebox{90}{ matrix}
		&\rotatebox{90}{ array}
		&\rotatebox{90}{ series}
		&\rotatebox{90}{ point/vector \hspace*{2mm}}
		\\
		\hline
		\sf Top10M 	& 12 & 4 & 1 & 5 & 1 & 9 & 3 & 5 & 0 & 7 & 11 & 9 & 3 & 4 & 9 & 2 & 5 \\
		\sf Top50P1 & 21 & 10 & 1 & 8 & 2 & 17 & 12 & 13 & 2 & 11 & 18 & 11 & 8 & 5 & 7 & 8 & 8 \\         
		\sf Top20P2 &  4 & 3 & 0 & 0 & 0 & 4 & 3 & 3 & 0 & 3 & 4 & 4 & 1 & 1 & 3 & 3 & 3 \\           
		\sf Random 	& 10 & 7 & 0 & 3 & 0 & 9 & 4 & 7 & 1 & 7 & 10 & 7 & 6 & 2 & 4 & 2 & 6 \\
		\hline
		\sf Total   & 47 & 24 & 2 & 16 & 3 & 39 & 22 & 28 & 3 & 28 & 43 & 31 & 18 & 12 & 23 & 15 & 22
	\end{tabular}
\end{table}

\begin{lstlisting}[language=Java,mathescape=true,escapeinside={(*}{*)},
%xleftmargin=.27\textwidth,
caption=Formula code in {\tt FloatCounter.java}~\cite{libgdx-floatcounter:file},label=floatcounterlisting:lst]
public void put (float value) {
    latest = value;
    total += value;
    count++;
    average = total / count;
    ...
\end{lstlisting}

Albeit recursive implementations are considered to be more elegant and closer to the mathematical specification, we found no recursively implemented mathematical formula in our sample. This may be due to the fact that recursive implementations suffer from performance penalties. Moreover, Grechanik et al.~\cite{Grechanik_et_al:esem10} analyzed 30.000 Java projects on SourceForge and found that less than 4\% of all methods were recursive.


\paragraph{Roles of Variables}
We categorized variables by the way they are initialized, read and changed in the formula code---i.e. the roles they play in the code. Table~\ref{coded:tab} lists the different roles of variables that we found in our examples. These different kinds of variable roles mostly refer to their use within a loop. First of all, it makes a difference whether a variable is read or written in the code. If a variable is only read, it is usually a parameter of a method, a field or a constant. 
More interesting are the write-accessed variables. Here, we distinguish 
accumulator variables, indexing variables and co-indexing variables.
Accumulator variables are used to accumulate a value over every iteration of a loop
and occur on the left side of an assignment expression. 
Indexing variables are usually incremented in each iteration of a loop and are mainly used to access data structures like arrays or collections, or used in an expression to generate a series of values. In {\tt for}-loops, the indexing variable is usually explicitly 
defined in the head of the loop. 

A co-indexing variable, like the name suggests, is a variable that indirectly depends on the value of the current indexing variable
and is possibly used in expressions in each iteration
of the loop. For example,
the variable {\tt pixel}  in the method {\tt analyze()} of the class {\tt HOGFeature.java} in the project {\verb+airbnb/aerosolve+} is a co-index of the index variable {\tt i} (see Listing ~\ref{lumlisting:lst}).
\begin{lstlisting}[float,language=Java,mathescape=true,escapeinside={(*}{*)},
%xleftmargin=.27\textwidth,
caption=SP-Formula code in {\tt HOGFeature.java}~\cite{aerosolve-hogfeature:file},label=lumlisting:lst]
// Compute sum of all channels per pixel
for (int y = 0; y < height; y++) {
  for (int x = 0; x < width; x++) {
    int pixel = image.getRGB(x, y);
    for (int i = 0; i < 3; i++) {
      lum[x][y] += pixel & 0xff;
      pixel = pixel >> 8;
    }
    lum[x][y] /= 3;
} }
\end{lstlisting}
\attic{Surprisingly,}
We only found few instances of co-indexing variables in our coded examples.
Finally, we also found temporary variables, which were used to split the computation of an
expression into several steps. We assume that temporary variables were often used to increase readability and to support debugging.  

Roles of variables have also been investigated in computer science education research~\cite{DBLP:conf/vl/Sajaniemi02}. Our roles of variables correspond to the ones
identified by Taherkhani et al.~\cite{Taherkhani2011} in their investigation of implementations
of sorting algorithms. Our term \textit{indexing variable} relates to their definition of a \textit{stepper}, our term \textit{co-indexing variable} to their definition of a \textit{follower}, and our term \textit{accumulation variable} to their understanding of a \textit{gatherer}. Finally, they denote constants or read-only variables as \textit{fixed values}. 



Often, valuable context information was encoded in the name of a variable, 
e.g. \texttt{C\_phi} ($C_\phi$) or \texttt{xbar} ($\overline{x}$). 
We assume that developers name variables or other artifacts like this when they program according to a visual representation of a mathematical formula. This way, a mental and partly visual connection between the program code and the underlying mathematical representation is established to increase the recognition value 
of the formula within the code. 


\paragraph{Comprehensibility}
The effort for reconstructing the formulas was influenced by several factors (see Table~\ref{coded:tab}). In many cases, the formulas could be reconstructed by inspecting the code fragment and without additional context information. In other cases, we had to inspect the code surrounding the actual denoted formula, because the code of the complete implementation can be fragmented in many lines of code, methods or even classes.\\
Furthermore, we recorded whether we were able to express the documented formula code in a mathematical notation (row \textit{expressible} in Table~\ref{coded:tab}). The few cases for which we could not reconstruct the underlying formula, but which were annotated by the developers to be implementations of some formula, are either indications of our limited domain knowledge or examples of how performance improvements obfuscate the program code.\\
A related property of the formula code is its proximity to the mathematical representation.
With 33 of 47 coded matches, the majority of the inspected formula code examples are quite close to their mathematical notation. 
When reconstructing the formulas, we often caught ourselves mixing mathematical notations with simple program code artifacts, e.g. function calls or array accesses. Therefore, we also 
coded whether a code fragment could be more intuitively described by a mixed representation of mathematical symbols and program code notation, even if we were able to reconstruct a purely mathematical representation as well. Figure~\ref{lumformula:fig} shows an example where we embedded code fragments into the mathematical representation since the bit operations \texttt{\&} and \texttt{{>}{>}} are programming specific and have no corresponding semantically equal symbols in maths. Altogether, for 18 of 47 
coded examples, we found that such a mixed form might be an appropriate alternative representation.
\paragraph{Mathematical Data Type}
Twelve of the 47 formula code examples implement matrix operations or calculations on matrices in general. Usually, matrices were implemented with the help of two-dimensional arrays either directly or indirectly through utility classes. 
But simply assuming a formula code example is implementing matrix calculations when detecting a two dimensional array is not always correct. Among other things, the correct sizes of the dimensions were necessary to consider a two dimensional array to be a matrix implementation. Furthermore, a two dimensional array in conjunction with a nested for-loop may not have been intended to be an actual matrix operation even though one could express it as such. 
Almost twice as many formula code examples (22 of 47) dealt with vector calculations or points in 2D or 3D space. 
In the coded examples, vector mathematics was implemented in three different ways:
by arrays (e.g. \texttt{v[0]}), 
by separate scalar variables for each dimension of a vector 
 (e.g. \texttt{v\_x}, mostly for 2D or 3D vectors), 
or by utility classes 
(e.g. \texttt{v.get(0)}).\\ 
Finally, 15 of 47 examples implemented a mathematical series with a single loop. In  mathematics, a series is an \textit{infinite} sum which of course is not implemented as such and thus outlines an important difference between program code representation and mathematical notation of the same formula.\\

\todo{OM: Auch eine Typisierung daraus ableiten? Bspw. Arithmetic, Summen, ...; sollten wir die 'Komplexitaet' auch bereits betrachten? -> Nein!}

\section{Derived Formula Code Patterns}
\label{patterns:sec}

In our exploratory study, we found many different structural aspects and other phenomena related to formula code. We also gained more insights in what developers annotated to be program code implementing a mathematical formula. Moreover, we were also able to derive patterns of typical formula code examples from our observations. In this section we will explain these in detail.

Since more than half of the examples in our qualitative study used {\tt for}-loops, we focus on loop-related patterns in the following. In particular, we look at those using \texttt{for} or \texttt{foreach} to implement a sum or product formula. In this section, we exemplary present a simple (non nested) \texttt{for}-loop and a nested \texttt{foreach}-loop pattern in an abstract notation including derived constraints for a transformation of the code to a formula representation in mathematical notation. 
Please note that arithmetic operations and function calls may occur in expressions within these formulas.
All patterns that we implemented in our detection tool 
to perform the case study in Section~\ref{casestudy:sec} can 
be described in the same manner. 

\subsection{Non-nested for loops}
A pattern of typical implementations of  sum and product formulas 
using a \texttt{for} loop is presented in Pattern~\ref{sumforpattern:pat}:

\newcommand{\blueOR}[0]{\color{blue}{$\!|\!$}}

\begin{pattern}[caption=Syntactic pattern for implementations of sum/product formulas using a \texttt{for}-loop,label=sumforpattern:pat]

for ( $index$ = $exp_1$; 
      $index$ < $exp_2$; 
      (*\color{blue}{[}*) $index$=$index$+1 (*\blueOR*) $index$+=1 (*\blueOR*) ++$index$ (*\blueOR*) $index$++ (*\color{blue}{]}*) ) { 
	$block_1$
	(*\color{blue}{[}*) $accu$ = $accu$ $op$ $exp_3$; (*\blueOR*) $accu$ $op$= $exp_3$; (*\color{blue}{]}*)
  	$block_2$
}
\end{pattern}	
In this pattern, we used the variable roles introduced in the previous section 
to name the meta variables. 
The meta variables $expr_i$ are placeholders for arbitrary expressions, $block_i$ 
for any possible other program code at those positions. This implies that the code implementing a formula can be embedded in  other program code and that only a small slice of code forms the actual formula code that we can represent in mathematical notation. In the above pattern, we increment the indexing variable by one in each iteration and make use of the comparison operator \textit{less than} in the loop conditional.
This is due to the fact that it is the most common utilization of a \texttt{for}-loop in this context, and a stepping of one is consistent with the semantics of a mathematical sum or product. Other loop-conditions or increments are conceivable but would result in a more complex mathematical representation albeit some, e.g. using the comparison operator \textit{less than or equal}, are trivial adjustments. In the patterns that we implemented, we therefore allowed other relational operators as well. Assuming that the value of the variable $accu$ is $accu_0$ before the first execution of the {\tt for}-loop, the mathematical formula that relates to the formula code pattern in Pattern~\ref{sumforpattern:pat}
is described by the following formula template:
\begin{eqnarray}
accu  = accu_0\ \ op \sum_{index=exp_1}^{exp_2-1} exp_3 & 
 {\color{blue}{\mbox{if $op\in \{ \texttt{+} , \texttt{-} \}$} } }\\
accu  = accu_0\ \ * \prod_{index=exp_1}^{exp_2-1} exp_3 & 
 {\color{blue}{\mbox{if $op\in \{ \texttt{*}  \}$} } } \\
accu  = accu_0\ \ * \prod_{index=exp_1}^{exp_2-1} \frac{1}{exp_3} & 
 {\color{blue}{\mbox{if $op\in \{ \texttt{/}  \}$} } }  
\end{eqnarray}
However, many program code fragments that match with the syntactic pattern may
not implement the formula described by the formula template, 
e.g. because they change the accumulator or
indexing variables in the code blocks or expressions. Hence, we extended the pattern 
with additional constraints to reduce the number of false positives.
To this end, we define the functions $vars()$ and $writes()$ 
which we will later use to formulate the constraints.
\begin{eqnarray}
\textbf{vars}(e) = \{ x \: |\  x\ \mbox{\rm occurs in\ }  e \}\\
\textbf{vars}(\{e_1, \dots, e_k\}) = \textbf{vars}(e_1) \cup \dots \cup \textbf{vars}(e_k)\\
\textbf{writes}(b)  = \{ v \: |\  v\verb+=+e\ \mbox{\rm occurs in\ } b \}\\
\textbf{writes}(\{b_1, \dots, b_k\}) = \textbf{writes}(b_1) \cup \dots \cup \textbf{writes}(b_k)
\end{eqnarray}

The first two constraints require that the accumulation variable $accu$ shall not occur in the expressions $expr_2$ and $expr_3$ and the indexing variable $index$ shall not occur in $expr_2$:
\begin{eqnarray}
accu \not\in \textbf{vars}(\{ exp_2, exp_3 )\\
index \not\in \textbf{vars}(exp_2)
\end{eqnarray}

We allow that the accumulation variable occurs in $expr_1$, because it will only be
evaluated once during the initialization of the indexing variable and at that time
it will have its initial value $accu_0$. It must not occur in $expr_2$, because then
it would occur on both sides of the equal sign in the formula.
The indexing variable must not occur in $expr_2$, because otherwise the upper bound of the sum or product would not be constant but reevaluated in each iteration.

The following three constraints require that any variable occurring in $expr_2$ and $expr_3$ as well as the variables $accu$ and $index$ will not be written within the body of the loop, 
i.e. that there are no assignment statements assigning new values to these variables:
\begin{eqnarray}
accu \not\in \textbf{writes}(\{ block_1, block_2\}) \\
index \not\in \textbf{writes}(\{ block_1, block_2, exp_3\}) \\
\textbf{vars}(\{exp_2, exp_3\}) \cap \textbf{writes}(\{ block_1, block_2, exp_2, exp_3\}) = \emptyset
\end{eqnarray}



Variables can not only be changed directly through assignments, but also indirectly through method calls in the body of the loop.
\attic{A called method might change the variables directly or indirectly.}
In this case, the method has a side effect:
Rountev \cite{Rountev2004} describes a side effect of a method as `(...) state changes that can be observed by code that invokes the method'. Thus, if all methods called in the body of the loop are side-effect free, we can be certain that they don't change the relevant variable values.
We did not add a constraint on side-effect-freeness, because it would require a full-fledged 
static program analysis.

\attic{
	
\todo{SD: Die Einschränkung auf F wurde nicht implementiert, oder?}
Variables can not only be changed directly through assignments, 
but also indirectly through method calls in the body of the loop. 
A called methods might change the variables directly or indirectly.
In this case, the method has a side effect.
Rountev \cite{Rountev2004} describes a side effect of a method as `(...) state changes that can be observed by code that invokes the method'. Thus, if all methods called in body of the loop are side-effect free, we can be certain that they don't change the relevant variable values.
We found that this requirement was too strong, because methods with side-effects (like {\small {\tt System.out.println()}}) might be called in the body of the loop, but would not have an effect on the computed formula. Thus, our last constraint only requires that all methods occurring in the expressions relevant for the formula are side-effect free:

\begin{equation}
\begin{array}{l}
F\ \mbox{\rm is the set of side-effect free methods}\\
\textbf{calls}(exp_i) \subseteq F
\end{array}
\end{equation}

}

\subsection{Nested loops}
We also defined patterns for nested loops, loops which iterate over arrays, 
and {\tt foreach}-loops which iterate over collections (including arrays). As an 
example, we present a pattern with two nested {\tt foreach}-loops below:

\begin{center}
\begin{pattern}[caption=Syntactic pattern for implementations of computing 
a vector of sums/products using two nested \texttt{foreach}-loops,label=arrayforeachpattern:lst]

for( $entry$ : $exp_1$ ) {
   $block_1$
   for( $elem$ : $exp_2$ ) {
      $block_2$
      (*\color{blue}{[}*) $entry$ = $entry$ $op$ $exp_3$ (*\color{blue}{|}*) $entry$ $op$= $exp_3$ (*\color{blue}{]}*)
      $block_3$
   }
   $block_4$
}      
\end{pattern}	
\end{center}
In the formula template related to this pattern, we use the mathematical notation
for an indexed family: 
\begin{eqnarray}
\left(entry\ \ op \!\!\!\!\sum_{elem \in exp_2} exp_3 \right)_{entry\in exp_1}  & 
 {\color{blue}{\mbox{if $op\in \{ \texttt{+} , \texttt{-} \}$} } }\\
\left(entry * \!\!\!\!\prod_{elem \in exp_2} exp_3 \right)_{entry\in exp_1} & 
 {\color{blue}{\mbox{if $op\in \{ \texttt{*}  \}$} } } \\
\left(entry * \!\!\!\!\prod_{elem \in exp_2} \frac{1}{exp_3} \right)_{entry\in exp_1} & 
 {\color{blue}{\mbox{if $op\in \{ \texttt{/}  \}$} } }  
\end{eqnarray}
To reduce the number of false positives we define the following constraints 
for the nested {\tt foreach} pattern:
\begin{eqnarray}
entry \not\in \textbf{writes}(\bigcup_{i=1}^{4} block_i) \\
elem \not\in \textbf{writes}(\{ block_2, block_3\}) \\
\textbf{vars}(exp_1) \cap \textbf{writes}(\bigcup_{i=1}^{4} block_i) = \emptyset \\
\textbf{vars}(\{exp_2, exp_3\}) \cap \textbf{writes}(\{ block_2, block_3, exp_3\}) = \emptyset 
\end{eqnarray}



\attic{
\todo{OM: later the following}
We also present an example of a syntactic pattern for vector addition as well as the computation of a vector dot product which both can be expressed mathematically equal through a sum formula.

\todo{OM: lower and upper bound? to fully implement a dot product, the lower bound has to be zero and the upper bound the length of the arrays to be certain that they are vectors!???!}
\begin{center}
\end{center}

\begin{displaymath}
accu  = \langle\vec{array_a},\vec{array_b}\rangle = \sum_{index=exp_1}^{exp_2-1} array_{a,index} * array_{b, index} 
\end{displaymath}

\todo{OM: what about the equality of definitions in math notations through equivalence transformations?}
\todo{OM: if in schleifenrumpf mit fallunterscheidung auch noch mit rein?}
}

\newcommand{\xsource}{ \mbox{\sf source} }
\newcommand{\xindex}{ \mbox{\sf index} }
\newcommand{\xsuffix}{ \mathit{suffix} }

\subsection{Vector arithmetics}

Although we did not use these patterns in our later analysis, we also defined patterns for vector addition and scalar product. We present a pattern for the 2D vector space which can easily be extended to more dimensions. Note that Pattern~\ref{sumforpattern:pat} would apply for scalar products as well if they are implemented using a \texttt{for}-loop. Furthermore, we did not observe any vector addition with more than three dimensions being implemented in the manner of the following pattern.

\begin{pattern}[caption=Syntactic patterns for implementations of vector addition and scalar product,label=vectorpattern:pat]

Scalar product:
$var$ = $exp_{1,1}$ * $exp_{2,1}$ + $exp_{1,2}$ * $exp_{2,2}$ ;
	
Vector addition:
$var_1$ = $exp_{1,1}$  + $exp_{2,1}$ ;
$var_2$ = $exp_{1,2}$  + $exp_{2,2}$ ;

\end{pattern}
\todo{OM, SD: align pattern3 horizontal, float, linebreaks=false not working} The formula template for these patterns uses typical vector notation:  

\begin{eqnarray}
var = \left\langle \begin{pmatrix} exp_{1,1} \\ exp_{1,2} \end{pmatrix} ,  \begin{pmatrix} exp_{2,1} \\ exp_{2,2} \end{pmatrix} \right\rangle
& \text{and} & 
 \begin{pmatrix} var_1 \\ var_2 \end{pmatrix} =
   \begin{pmatrix} exp_{1,1} \\ exp_{1,2} \end{pmatrix} + \begin{pmatrix} exp_{2,1} \\ exp_{2,2} \end{pmatrix}
\end{eqnarray}
The above patterns would match with far too many assignments in the source code, thus
we add the following constraints.
Our constraints are based on the observation that certain suffixes occur often: \verb+.x+; \verb+getX()+ or \verb+[0]+ and that certain naming conventions are used, e.g. \verb+sx+ or \verb+s1+, to access components of a vector. First, we define the following auxiliary functions:
\begin{eqnarray}
\xsource(e) = e', \mbox{\rm if } e=e'{\tt .}s \mbox{\rm\ and\ } s \in \{ {\tt x}, {\tt y}, {\tt getX()}, {\tt getY()}, {\tt get(0)}, {\tt get(1)} \} \mbox{\rm \ or\ } e=e'{\tt [}i{\tt ]}\\
\xsource(e) = p, \mbox{\rm if } e=ps \mbox{\rm \ \text{is a variable name with prefix $p$ and suffix $s$}} \in \{ {\tt x}, {\tt y}, 0, 1\}\\
\xindex(e) = 0, \begin{array}[t]{l}\mbox{\rm if } e=e'{\tt .}s \mbox{\rm\ and\ } s \in \{ {\tt x}, {\tt getX()}, {\tt get(0)} \}\\
\mbox{\rm\ or\ } e=e'[0] \mbox{\rm\ or} \mbox{\rm\ $e$ \text{is a variable name with suffix $s$}} \in \{ {\tt x}, 0 \}
\end{array}\\
\xindex(e) = 1, \begin{array}[t]{l} \mbox{\rm if } e=e'{\tt .}s \mbox{\rm\ and\ } s \in \{ {\tt y}, {\tt getY()},  {\tt get(1)} \}\\ \mbox{\rm\ or\ } e=e'[1] \mbox{\rm\ or} \mbox{\rm\ \text{$e$ is a variable name with suffix $s$}}\in \{ {\tt y}, 1 \}
\end{array}
\end{eqnarray}
Now we can define the following constraints:
\begin{eqnarray}
\xsource(e_{1,1}) = \xsource(e_{1,2}) \mbox{\rm\ and\ }\xsource(e_{2,1}) = \xsource(e_{2,2})\\
\xindex(e_{1,1}) = \xindex(e_{2,1})  \mbox{\rm\ and\ }\xindex(e_{1,2}) = \xindex(e_{2,2})
\end{eqnarray}
For the vector addition we also require:
\begin{eqnarray}
\xindex(var_{1}) = \xindex(e_{1,1}) = \xindex(e_{2,1})\\
\xindex(var_{2}) = \xindex(e_{1,2}) = \xindex(e_{2,2}) \\
\xsource(var_1) = \xsource(var_2) 
\end{eqnarray}
Now the formula templates can be rewritten as:
\begin{eqnarray}
var = \left\langle\, \xsource(exp_{1,1})  , \xsource(exp_{2,1}) \,\right\rangle\\  
\xsource(var_1) = \xsource(exp_{1,1}) + \xsource(exp_{2,1}) 
\end{eqnarray}

\newcommand{\xvars}{ \mbox{\sf vars} }
\newcommand{\xwrites}{ \mbox{\sf writes} }
\newcommand{\xcalls}{ \mbox{\sf calls} }

Note that the presented patterns with the given constraints are only a heuristics to be able to find candidates for formula code, i.e. instances of the pattern within Java software projects. The patterns neither define necessary nor sufficient conditions of program code implementing, for example, a sum or product formula. 

\section{SP-Formula Code on GitHub \attic{:A Case Study}}
\label{casestudy:sec}
To answer research question RQ2, we performed a 
quantitative study on two different samples of open source Java projects on GitHub. To automate the search for formula code, in particular SP-formula code, we developed a detection tool which employs refined variations of the patterns introduced in Section~\ref{patterns:sec}. First, we present some detail on the SP-formula code detection tool, in particular on the patterns it can detect and its evaluation in terms of precision and recall. Next, we introduce a sample of engineered Java software projects
of arbitrary topic (see GitHub \textit{topics}~\cite{GitHubHelpTopic2018}) and present the results obtained by our tool for this sample. We also look at the application domains of the projects with highest SP-formula code densities. Thereafter, we describe another sample consisting solely of Java projects with topic \textit{scientific-computing} and discuss the 
results computed by our tool for this sample.

\attic{ We then compare the results obtained by our tool for both samples followed by a discussion to answer the research questions. }

\todo{OM: We didn't implement all of the patterns since vector/matrix code is a different kind/type of formula code?}

\subsection{Pattern-Based SP-Formula Code Detection Tool}
\label{tool:sec}
For our study, we used a shell script that clones each project from a given list of GitHub projects and checks out their configured default branch. 
The detection tool reads all Java files of these projects one after another (comments in the source code are removed) and searches for all patterns in parallel to exploit multiple CPU cores.
Each pattern is implemented as a compiled regular expression. Although the regular expressions cannot match every syntactic pattern of the Java programming language, the approach performed pretty well for our purposes as it also captures and preprocesses pattern-specific elements like variable roles, such that the constraints associated with each pattern can be tested. The tool tests the nested SP-formula code patterns first and then the non-nested SP-Formula-code patterns (see Table~\ref{patterntypes:tab}). Every match of the non-nested patterns will be checked for intersection with all matches of the nested patterns. If an intersection exists, the respective non-nested match will be discarded. Matched code fragments that satisfy the constraints are attached to the output which is stored in form of a CSV file. This file lists for each entry, the line numbers of the start and end of the match, the source code of the code fragments, the inferred formula in mathematical notation as well as other detailed information about the source like project name, file name and GitHub path.
The inferred formula is basically an instance of our formula templates and is stored in the file both in a textual representation as well 
as in MathML.

Listing~\ref{regularexp:lst} shows the regular expression 
for Pattern~\ref{sumforpattern:pat} (all pattern implementations are available in the supplementary material~\cite{moseler_oliver_2018_1252324}). Note that the regular expression 
is built using String constants like {\tt VAR} and {\tt EXP} which themselves contain regular expressions.

\setcounter{lstlisting}{2}
\begin{lstlisting}[float,language=Java,mathescape=true,escapeinside={(*}{*)},basicstyle=\ttfamily,caption=Regular expression for non-nested \texttt{for}-loop implementing a sum/procuct  (Pattern~\ref{sumforpattern:pat}),label=regularexp:lst]
 Pattern.compile(
   (*\color{blue}{// head of loop}*)
   "(?<lineOuterFor>for"+b+"\\((?:"+b+DT+b2+")?"+
   "(?<ind0>"+VAR+")"+b+"="+b+"(?<exp00>"+EXP+")"+
   b+";"+b+"\\k<ind0>"+b+"(?<relOp0>"+REL_OP+")"+b+
   "(?<exp10>"+EXP+")"+b+";"+ITER0_INCREASE+"\\))\\s*+"
   (*\color{blue}{// body in brackets}*)				
   + "(?:\\{\\s*?" + 
   "(?<blockFiOu>"+BLOCKR+"\\s++)" + 
   "(?<lineAssiA>"+ACCUA_ASSIGNMENT+";)\\s*+" +
   "(?<blockSeOu>"+BLOCKP+")\\s*+" +
   "\\}"
   (*\color{blue}{// no brackets}*)							
   + "|" +
   "(?<lineAssiB>"+ACCUB_ASSIGNMENT+";))"
 );
\end{lstlisting}

\attic{Given the diversity of the formulas that we found, we decided to focus on sum and product formulas for our quantitative analysis, because they are non-trivial and we expected them to occur quite frequently.Thus,} 
We implemented the 10 patterns listed in Table~\ref{patterntypes:tab} to detect SP-formula code. On average, the regular expressions for non-nested loops consist of 8 lines of code without comments, those for nested loops of 28 lines.



\attic{
	\begin{description}[align=left, font=\normalfont\normalcolor\sffamily] 
		\item [FIS:] non-nested {\tt for}-loop for sum/product
		\item [FES:] non-nested {\tt foreach}-loop for sum/product
		\item [FIA:] non-nested {\tt for}-loop for arrays
		\item [FEC:] non-nested {\tt foreach}-loop for arrays/collections
		\item [NFISS:] nested {\tt for}-loops for sum/product of sums/products
		\item [NFESS:] nested {\tt foreach}-loops for sum/product of sums/products
		\item [NFIAS:] nested {\tt for}-loops for array of products/sums
		\item [NFECS:] nested {\tt foreach}-loops for array/collection of products/sums
		\item [NFIAA:] nested {\tt for}-loops for array of arrays
		\item [NFECC:] nested {\tt foreach}-loops for array/collection of arrays/collections
	\end{description}
}

\begin{table}[htb]
	\centering
	\caption{Patterns implemented by the tool~\label{patterntypes:tab}}

	\begin{tabular}{cll}
		\hline
		\multirow{4}{*}{\rotatebox{90}{\parbox{1.7cm}{\center \footnotesize Patterns for\\ non-nested loops}}} & FIS & {\tt for}-loop for sum/product \\
		&FES & {\tt foreach}-loop for sum/product\\
		&FIA & {\tt for}-loop for arrays\\
		&FEC & {\tt foreach}-loop for arrays/collections\\
		\hline\hline
		\multirow{6}{*}{\rotatebox{90}{\parbox{2cm}{\center \footnotesize Patterns for nested loops}}} & NFISS & {\tt for}-loops for sum/product of sums/products\\
		&NFESS & {\tt foreach}-loops for sum/product of sums/products\\
		&NFIAS & {\tt for}-loops for array of products/sums\\
		&NFECS & {\tt foreach}-loops for array/collection of products/sums\\
		&NFIAA & {\tt for}-loops for array of arrays\\
		&NFECC & {\tt foreach}-loops for array/collection of arrays/collections\\
		\hline
	\end{tabular}
\end{table}

\paragraph{Evaluation}\label{evaluation:sec}


To evaluate our approach to detect SP-formula code in software repositories we applied our tool with the patterns listed in Table~\ref{patterntypes:tab} to random samples of Java files on GitHub. As validation metrics we use recall and precision:
\begin{itemize}
	\item Recall: How many of the SP-formula code fragments in a sample are automatically detected?
	\item Precision: To what extent are the detected SP-formula code fragments correct?
\end{itemize}

More precisely, let $F$ be the set of all formula code fragments in the sample, and $D$ be the detected formula code fragments. Then the recall is the number of correctly found formula code fragments divided by all correct formula code fragments, i.e., ${\sf recall} = \frac{|D\cap F|}{|F|}$, and the precision is the number of correctly found formula code fragments divided by all found formula code fragments, i.e., ${\sf precision} = \frac{|D\cap F|}{|D|}$.
\attic{
	\todo{SD: Das stimmt ja so nicht, da wir in der Studie watcher $>$ 10 nehmen:
		Since there are no sufficient studies about the quality, properties and characteristics of the data available on GitHub, there do not exist any guidelines to chose adequate samples
	}
}
Based on the GBQ GitHub and GHTorrent dataset, we retrieved a list of 21,052,682 Java filenames (including full path information) on GitHub. The list does not include any forked repositories and duplicates, i.e. files with the same hash value.
To draw random samples from the above data set we used the statistical programming language R and its uniform distributed function `runif'. To measure the recall of our approach we used a sample of 1,000 Java files, to measure its precision we used a sample of 10,000 files. 
Since the GBQ GitHub and GHTorrent datasets are off-line mirrors of GitHub metadata, each sample contained names of files 
which have already been moved, removed or renamed on GitHub. 
Thus, we could not download the source code of all files leading to a sample size of 878 and 8,960, respectively.
\paragraph{Oracle} \label{para:oracle}
For computing the recall, we created an oracle data set by manually inspecting the complete sample of 878 files and annotating the formula code fragments that we found. We annotated the code fragments in a group discussion in order to reduce subjective biases.
To annotate the fragments, we enclosed them in XML-tags which can be nested. The choice of 
the tags {\verb+<SimpleNestedLoop>+}, {\verb+<DoubleNestedLoop>+},
{\verb+<SimpleArithmetic>+}, 
{\verb+<Matrix>+}, and {\verb+<Vector>+} is based on the results of our keyword-based search. Finally,
we manually found and annotated 145 formula code fragments in 53 
of the 878 files (6.04\%). These formula code fragments 
made up 1,064 lines of 142,419 total lines of code (0.75\%).
These annotated formula code fragments contained SP-formula code
in 110 cases (75.86\%). 
Almost all remaining cases were annotated with 
{\verb+<SimpleArithmetic>+} which we used to describe simple, but non-trivial arithmetic expressions. The mathematical representation of those expressions may have some added value compared to 
the representation in program code, 
e.g. $\sqrt{5}$ instead of {\verb+Math.sqrt(5)+}.
Matrices and vectors were only used in 4 lines outside of loops.

\paragraph{Recall}
Our tool found 34 of 110 SP-formula code fragments in the oracle data set. Thus, 
the recall is 30.91\%. 
In cases where both an inner and outer loop each represent an SP-Formula code fragment, the tool would only consider the outer one. However, in our evaluation this case never happened.

\attic{
	\begin{figure}[ht]
		\centering
		\caption{Recall of the Formula Code Mining Tool\label{toolrecall:fig}}
		\includegraphics[scale=1]{./images/Recall}
	\end{figure}
}


\paragraph{Precision}
To measure the precision we applied our tool to the 
8,960 Java files of the second, non-annotated sample. For each detected
formula code fragment, we manually checked whether the matched
code fragment implemented an SP-formula. We also recorded whether the inferred formula covered the matched code fragment completely or only a part of it.

Our tool found 181 matches. All of these matches contained SP-formula code. 153 code fragments got completely specified by the inferred formula in mathematical notation. 
For 23 matches the tool inferred a correct formula that, however, did not describe the whole code excerpt. 
Only in 5 cases we found that the inferred mathematical formula was
inadequate or wrong (2.76\%). 
Thus,\attic{ as shown in Figure~\ref{toolprecision:fig},}
if we only take into account whether
the match contained formula code, the precision was 100\%. 
If we require that the inferred formula is correct, the precision
was 97.23\%, and if we require that the inferred formula is correct 
and completely describes the effect of the code matched, the precision
was 84.53\%.

\attic{
	\begin{figure}[ht]
		\centering
		\caption{Precision of the SP-Formula-Code Detection\label{toolprecision:fig}}
		\includegraphics[scale=1]{./images/Precision}
	\end{figure}
}

\paragraph{Formula code density}
\label{fcdensity:para} To quantify how often formula
code occurs in real world software (RQ2), we define two different 
measures for the formula code density---one
based on lines of codes $\rho_{LOC}$
and one based on number of unique files $\rho_{files}$:
\begin{eqnarray} \label{fcdensityfiles:equ}
\rho_{files} = \frac{\mbox{\#files containing formula code}}{\mbox{\#all scanned files}}\\
\rho_{LOC} = \frac{\mbox{\#lines with formula code in all scanned files}}{\mbox{\#lines of all scanned files}}\label{fcdensityloc:equ}
\end{eqnarray}

\attic{
	Figure~\ref{oracleFCdensity:fig} shows the formula code 
	density in our oracle data set, i.e. based on all manually 
	identified and annotated formula code fragments.
}


The formula code densities in our oracle data set, i.e. based on all manually identified and annotated formula code fragments, were $\rho_{files}=\frac{53}{878}= 6.04\%$ and $\rho_{LOC}=\frac{1,064}{142,419}= 0.76\%$. In other words, in our oracle data set on average one of 130 lines of code was part of a code fragment which we annotated as formula code.

\attic{
	\begin{figure}[ht]
		\centering
		\caption{Formula Code Density in Oracle Data Set\label{oracleFCdensity:fig}}
		\includegraphics[scale=.9]{./images/oracleFormulaDensity}
	\end{figure}
}

\subsection{SP-Formula Code in engineered Java software projects on GitHub}
In the following, we introduce the examined sample of randomly chosen open source Java stargazer projects, present results of our detection tool for this sample and look at the application domains of the SP-formula code-rich projects.

\attic{
\begin{description}[font=\normalfont\normalcolor\sffamily]
\item [TOP50P1:] First, we applied the tool to sample {\sf Top50P1} from 
Section~\ref{keyword:sec} to compare the number of SP-formula code fragments found 
by our  tool to  the number of SP-formula code fragments found based on keywords and 
manual inspection.
\item [Random2:] Second, we generated a completely new sample of 1000 randomly chosen non-forked open source Java projects from GitHub excluding the already examined projects from our preliminary qualitative study. We utilized Google Big Query to generate a list of
non-forked OSS Java projects. The list contained 255,561 projects (as of January, 25th 2017) and we used the statistical programming language R to generate the random samples. 
\item [Stargazers:] Third, we applied the stargazers-based classifier with 
threshold 10 which according to a recent study by Munaiah et al.~\cite{Munaiah2017} 
has high precision (97\%) and a for our purposes sufficient recall (32\%) to predict whether a GitHub project is an engineered software project.
We used the list of non-forked open source Java projects mentioned above, 
which we filtered by watcher (stargazers) count greater 
than 10. The filtered list contained 28,139 projects, from which we randomly drew 100 
for the new sample, this time excluding both the already examined projects of our preliminary qualitative study and those of the previous sample {\sf Random2}.
\end{description}
}

\paragraph{\sf Stargazers}
We used the stargazers-based classifier approach with threshold 10 which according to a recent study by Munaiah et al.~\cite{Munaiah2017} has high precision (97\%) and a reasonable recall (32\%) to predict whether a GitHub project is an engineered software project and thus is sufficient for our purposes. First, we generated a sample of randomly chosen non-forked open source Java projects from GitHub excluding the already examined projects from our preliminary qualitative study. We utilized the GBQ GHTorrent dataset to retrieve the initial project list which contained 255,561 projects (as of January, 8th 2019). Next, we filtered the list by watcher (\textsf{Stargazers}) count greater than 10. The filtered list contained 28,139 projects, from which we randomly drew 1000 with the help of the statistical programming language R.
The results of applying our SP-formula code detection tool to the \textsf{Stargazers} sample, the \textsf{SciC} (scientific computing) sample (see Subsection~\ref{github-scic:subsec}) as well as aggregated results are shown in Table~\ref{toolapplsamples:tab}. It lists the total number of projects investigated in each sample (\#projects), the number of projects containing any kind of Java code at all (\#nonempty), the number of projects which contained any SP-formula code detected by the tool (\#fc projects), the total number of Java files within the complete sample (\#files), the total number of files which contained any SP-formula code detected by the tool (\#fc files), the total number of lines of Java code in the complete sample (LOC), the total number of detected lines of SP-formula code (LOFC), the total number of matches found by the tool in the complete sample (\#matches) as well as code densities for actually detected SP-formula code $\rho^{SP}_{files}$ and $\rho^{SP}_{LOC}$ based on the definitions of general formula code densities in Equation \ref{fcdensityfiles:equ} and \ref{fcdensityloc:equ}, respectively.

\todo{R2: Assuming, that the ratio of SP-formula code that is detectable vs. non-detectable by our tool is the same in the samples.}

Assuming, that the distribution of SP-formula code in the samples is the same as in the oracle, we can use the recall of 31\% determined in Section~\ref{evaluation:sec} to compute a rough estimation of the real SP-formula code densities $\widetilde{\rho}^{\,SP}_{\,LOC} = \frac{1}{\sf recall}\, \rho^{SP}_{LOC}$,
resp. $\widetilde{\rho}^{\,SP}_{\,files} = \frac{1}{\sf recall}\, \rho^{SP}_{files}$.

\begin{table}
\caption{Results of the SP-formula code detection tool for each sample~\label{toolapplsamples:tab}}
\centering
\begin{tabular}{rrr||r}
  Sample & \textsf{Stargazers} & \textsf{SciC} & Sum\\
  \hline
  {\sf \#projects} & 1000 & 14 & 1014 \\
  {\sf \#nonempty} & 949 & 14 & 963 \\
  {\sf \#fc projects} & 266 & 11 & 277 \\
  {\sf \#files} & 199,457 & 4050 & 203,507 \\
  {\sf \#fc files} & 1713 & 199 & 1,912  \\
  {\sf LOC} & 30,275,938 & 548,976 & 30,824,914 \\
  {\sf LOFC} & 13,094 & 1,794 & 14,888 \\
  {\sf \#matches} & 2,858 & 515 & 3,373  \\
  {\sf $\rho^{SP}_{files}$} & $0.85\%$ & $4,91\%$ & $0.94\%$ \\
  {\sf $\rho^{SP}_{LOC}$} & $0.043\%$ & $0.32\%$ & $0.048\%$ \\
  {\sf $\widetilde{\rho}^{\,SP}_{\,files}$} & $2.74\%$ & $15.84\%$ & $3,03\%$ \\
  {\sf $\widetilde{\rho}^{\,SP}_{\,LOC}$} & $0.14\%$ & $1,03\%$ & $0.15\%$ \\
  \hline
\end{tabular}
\end{table}

LOC was computed with the Unix command \textsf{cloc} (version 1.74) and does neither include comments nor performs a uniqueness test on files. The same holds for LOFC computed by our tool, since it removes comments before scanning the files and processes every file in the project. Further statistical analyses were done with R. In total, we scanned 199,457 Java files from 949 open source Java projects of arbitrary topic available on GitHub using the patterns listed in Table~\ref{patterntypes:tab}. Below, we present the results for this sample.

\attic{

\paragraph{Results for \textsf{TOP50P1}}
The tool yields 1,858 SP-formula code matches in 40 of the 49 projects in
this sample. Thus, 81\% of the projects in this sample contain at least 
one file which itself contains SP-formula code.
Considering the number of files within these projects, SP-formula code was
detected in 1014 of the 115,936 files in this sample, thus 
the density of detected SP-formula code based on files $\rho^{SP}_{files}$ is 0.87\%.
Furthermore, 7,084 of 10,899,067 total lines of Java code contained detected
SP-formula code, hence the density of detected SP-formula code based on lines of code $\rho^{SP}_{LOC}$ is 0.065\%. 
Practically speaking, on average the tool detected SP-formula code in one of
114 files resp. in one of 1538 lines of code.
\attic{
Assuming, that the distribution of SP-formula code in the sample is the
same as in the oracle, using the recall of 23\% determined
in Section~\ref{evaluation:sec} we computed a rough estimation
of the real number of lines with SP-formula code $\widetilde{LOFC}
= \frac{6,879}{23\%} \approx 29,900$, and thus a rough
estimation of the SP-formula code density $\widetilde{\rho}_{LOC}
= \frac{\rho_{LOC}}{23\%}\approx 0.29\%$, or in other words,
we expect that one of 340 lines of code implements a sum or product formula.
}

Figure~\ref{fcmatchesinprojects:fig} shows the variation of the total number of SP-formula code matches for each sample. In the box plot for this sample, we see that there are a few outliers: single projects which contain several hundreds up to 430 detected SP-formula code fragments, i.e. many more matches compared to all other projects in this sample. 

}

\attic{

\paragraph{Results for \textsf{Random2}}
In the second sample, the tool detected 1248 SP-formula code fragments in 204 of the 992 scanned projects (20.5\%) resp. in 774 of the 79,371 Java files. A total of 5,736 lines of the
8,055,976 total lines of code in this sample were part of a detected SP-formula code fragments.
Thus, the densities of SP-formula code in this sample are $\rho^{SP}_{files}=0.98\%$ and $\rho^{SP}_{LOC}=0.071\%$.
\attic{Again, we can compute rough estimations
of the real number of lines with SP-formula code $\widetilde{LOFC}= \frac{697}{23\%} \approx 3030$, 
and the SP-formula code density $\widetilde{\rho}_{LOC}
= \frac{0.073\%}{23\%}\approx 0.32\%$, which is 
very close to the value of the previous sample.  
}

The box plot for this sample in Figure~\ref{fcmatchesinprojects:fig} contains four outliers: in particular, projects with 111, 86, 61 and 54 matches. The remaining 936 matches are spread over 200 projects.


}

\begin{figure}
	\centering
	\includegraphics[scale=0.8]{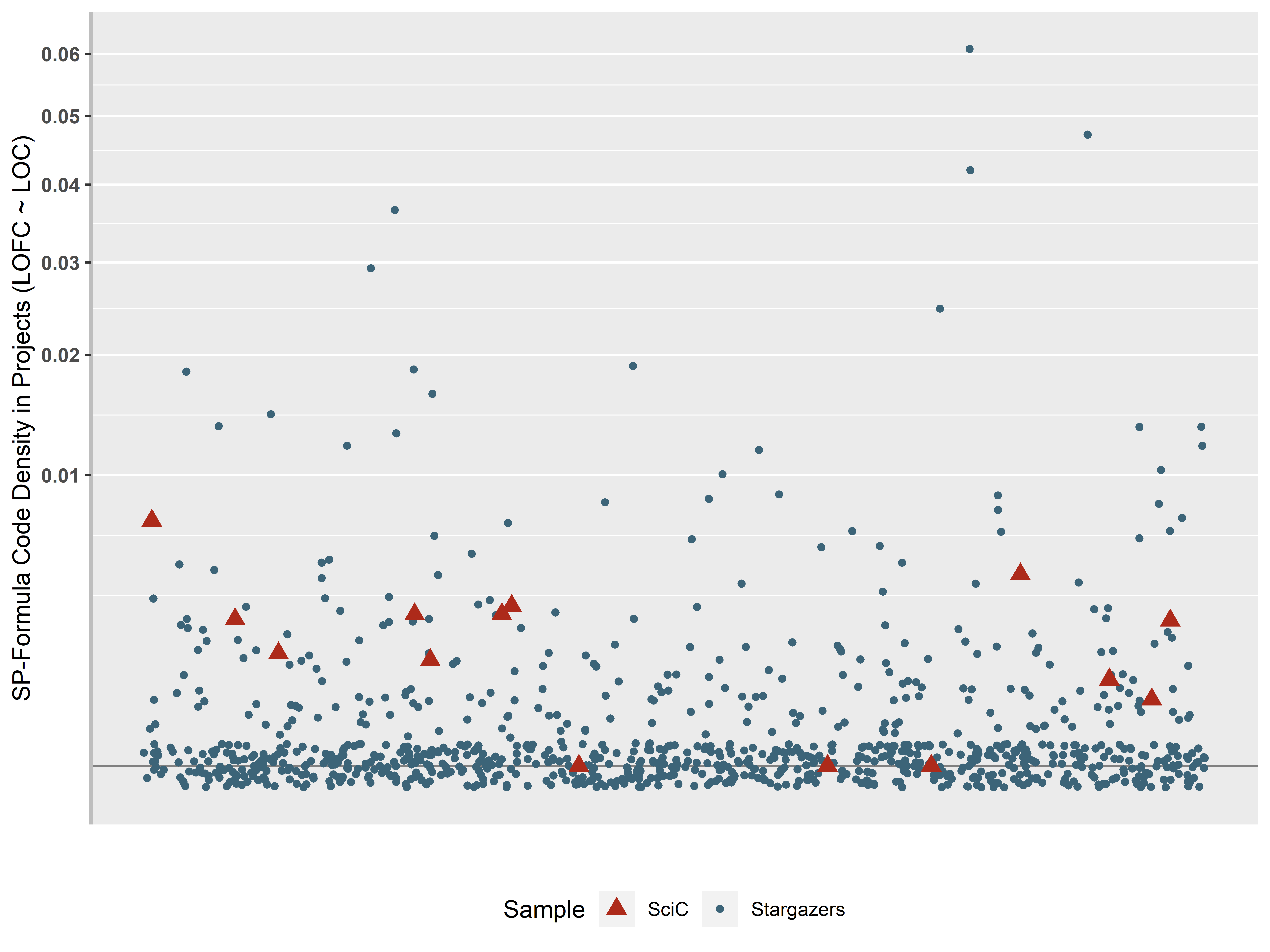}
	\caption{Jittered scatter plot of SP-formula code densities $ \rho^{SP}_{LOC} $ across all projects for each sample.}
	\label{fcdensityinprojects:fig}
\end{figure}

\paragraph{Results for \textsf{Stargazers}}
Our detection tool yielded 2,858 SP-formula code matches in 266 of 949 projects (28\%) respectively in 1713 of 199,457 Java files in this sample. The detected SP-formula code was spread over 13,094 lines of 30,275,938 total lines of Java code. Thus, the densities of SP-formula code in this sample are $\rho^{SP}_{files}=0.85\%$ and $\rho^{SP}_{LOC}=0.043\%$. Practically speaking, on average the tool detected SP-formula code in one of 117 files respectively in one of 2325 lines of code.
\attic{The box plot in Figure~\ref{fcmatchesinprojects:fig:box} reveals that the majority (2436) of the matches are spread over only a few projects (82, with number of matches greater or equal to 7) with one project already containing 9.7\%, i.e. 227, of all detected matches. The remaining 422 SP-formula code occurrences are spread over another 184 projects in the sample whereas 683 projects did not contain any match at all. Furthermore, the interquartile range is small and while the median value is 0, the mean value of 3.565 is surprisingly high (see Figure~\ref{fcmatchesinprojects:fig:box} and~\ref{fcmatchesinprojects:fig:scatter}). 
}
As can be seen in Figure~\ref{patterndens:fig}, every pattern derived from our preliminary study occurred in the \textsf{Stargazers} sample, which confirms their relevance. Nevertheless,
the non-nested loop patterns \textsf{FIS}, \textsf{FES} and \textsf{FIA} are more common than  the nested ones. Compared to the other non-nested loop patterns, the pattern \textsf{FEA} describing a foreach-loop that traverses an array sticks out, because it is rarely found.\\
\attic{The overall results tell us so far, that the probability to discover a project containing SP-formula code by randomly choosing a stargazing project from GitHub is very low. 
Due to a missing relation to the size of a project,
}
Considering the absolute number of matches is not sufficient to tell if a project contains much SP-formula code. Therefore, we further investigate the SP-formula code density, $ \rho^{SP}_{LOC} $ introduced earlier in this section to identify SP-formula code rich projects. A corresponding scatter plot is presented in Figure~\ref{fcdensityinprojects:fig}.
It becomes apparent that only few projects (18, with a density $ \rho^{SP}_{LOC} $ greater or equal to 0.01) have comparatively high formula code densities, i.e. at least 1 of 100 lines of code is part of SP-formula code. Table~\ref{highdensitystargazersprojects:tab} shows the complete list of these projects. 

\attic{
\begin{figure}
	\centering
	\includegraphics{images/formula-code-matches-in-stargazers-per_project-boxplot.png}
	\caption{Absolute number of SP-formula code matches across all projects}
	\label{fcmatchesinprojects:fig:box}
\end{figure}

\begin{figure}
	\centering
	\includegraphics[scale=0.8]{images/formula-code-matches-in-stargazers-scic-per-project-scatterplot.png}
	\caption{Jittered scatter plot of absolute number of SP-formula code matches across all projects }
	\label{fcmatchesinprojects:fig:scatter}
\end{figure}
}


\begin{table}
	\caption{Top 18 projects of the \textsf{Stargazers} sample having high (greater or equal to 0.01) SP-formula code densities ordered by {\sf $\rho^{SP}_{LOC}$} including their respective coded application domain.}
	\label{highdensitystargazersprojects:tab}
	\centering
	\begin{tabular}{l|ccc|r}
		Project & LOFC & LOC & {\sf $\rho^{SP}_{LOC}$} & Application Domain\\
		\hline 
		grunka/Fortuna & 87 & 1476 & 5.8\% & Cryptography\\
		hfut-dmic/ContentExtractor & 16 & 355 & 4.5\% & Information Retrieval \\
		eljefe6a/UnoExample & 3 & 75 & 4.0\% & Programming Model \\
		radzio/AndroidOggStreamPlayer &	272 & 7068 & 3.8\% & Signal Processing \\
		brendano/myutil & 319 & 10930 &	2.9\% & Statistics \\
		InfiniteSearchSpace/Automata-Gen-3 & 100 & 3694 & 2.7\% & Simulation \\
		DASAR/Minim-Android & 186 & 9604 & 1.9\% & Signal Processing \\
		hanks/Natural-Language-Processing & 79 & 4358 & 1.8\% & Natural Language Processing \\
		liuxang/LivePublisher & 11 & 636 & 1.7\% & Signal Processing \\
		sudohippie/throttle & 12 & 730 & 1.6\% & Computer Networks \\
		jlmd/SimpleNeuralNetwork & 6 & 423 & 1.4\% & Machine Learning \\
		ozelentok/CodingBat-Soultions & 36 & 2593 & 1.3\% & Programming Language Practice \\
		aliHafizji/CreditCardEditText & 4 & 307 & 1.3\% & Mobile User Interface \\
		quiqueqs-BabushkaText & 4 & 318 & 1.2\% & Mobile User Interface \\
		jroper-play-promise-presentation & 3 & 239 & 1.2\% & Programming Model \\
		jestan/netty-perf & 8 & 669 & 1.1\% & Computer Networks \\
		chipKIT32/chipKIT32-MAX & 674 & 62409 & 1.0\% & Micro-Controller \\
		martijnvdwoude/recycler-view-merge-adapter &  3 & 281 & 1.0\% & Mobile User Interface \\
		\hline
	\end{tabular}
\end{table}

The 18 projects with high SP-formula code density form the basis for a coding of the respective application domains and thus towards further investigations concerning RQ2.

\paragraph{Application Domains}
As there was no sufficiently exact and efficient way to automatically determine the application domain of GitHub projects, we manually inspected the web sites of the projects to extract this information.  
While GitHub offers the feature \textit{topics}~\cite{GitHubHelpTopic2018}, 
\attic{ which allows users to tag repositories to add meta information like intended purpose, subject area or affinity groups to their project,}
this feature is not sufficiently informative for the majority of projects examined in this work. 

As it was not possible with reasonable effort to determine the application areas for all projects included in our sample, we decided to take a closer look at the application domains of the projects having a high SP-formula code density, more precisely, the top 18 projects of the \textsf{Stargazers} sample in terms of $\rho^{SP}_{LOC}$ as presented in Table~\ref{highdensitystargazersprojects:tab}.
Again we applied open and axial coding to determine the application domains based on the descriptions of these projects on GitHub and names of code artifact, such as classes and packages. In total, we performed two iterations involving three of the four authors, where we abstracted from more specific to more generic categories. The resulting application domains are also shown in Table~\ref{highdensitystargazersprojects:tab}.

The results of the application domain coding do partially coincide with our expectations. One SP-formula code dense project is a practice project, where features of the Java programming language are exercised. Another two projects highlight a certain programming model. Furthermore, three projects are concerned with mobile user interface elements. These projects form outliers in the sense that their high formula code density is due to their low number of lines of code, because they only contain a single SP-formula code match (see Table~\ref{highdensitystargazersprojects:tab}).\\ 
The other twelve projects draw a different application domain picture. We labeled three projects with \textsf{Signal Processing} as they dealt with either audio or video encoding, respectively transfer. Besides that, we assigned the labels \textsf{Computer Networks}, \textsf{Statistics}, \textsf{Information Retrieval}, \textsf{Neural Networks}, \textsf{Natural Language Processing}, \textsf{Micro-Controller}, \textsf{Simulation} and \textsf{Cryptography} 
to characterize the application domains of the remaining SP-formula code dense projects. For each of the labels assigned to these twelve projects, we can associate the application of mathematical formulas. Surprisingly, neither domains like computer graphics and games, nor chemistry and physics, which one typically subsumes by the term scientific-computing, occurred among the projects with the highest SP-formula-code density in our sample. Nonetheless, the majority of the assigned labels seem related to scientific-computing. Thus, projects of this application domain seem promising in terms of frequency of SP-formula code and we investigate a thematically more focused sample as described in the following.

\todo{SD: Wuerde den Satz streichen:  We can only speculate that these kinds of projects neither are open source, written in Java, been sampled in our study nor be representative application domains in terms of SP-formula code according to our definitions. 
}

\subsection{SP-Formula Code in Scientific-Computing Java Projects on GitHub}
\label{github-scic:subsec}
Due to the unexpected distribution of application domains among the projects with the highest SP-formula-code density, we drew another sample solely consisting of Java repositories with topic \textit{scientific-computing}.

\attic{
\begin{description}[font=\normalfont\normalcolor\sffamily]
\item [\sf SciC:] To generate this topic-specific sample, we used the search term \texttt{language:Java topic:scientific-computing} to query Java repositories directly from the GitHub website. The ten resulting repositories (as of May, 18th 2018) do neither intersect with the already examined projects of our preliminary qualitative study nor with one of the samples {\sf Random2} or {\sf Stargazers}.
\end{description}
}

\paragraph{\sf SciC}
To generate this topic-specific sample directly from the GitHub website, 
we used the search term \texttt{language:Java topic:scientific-computing} to query Java repositories. The 14 resulting repositories (as of January, 9th 2019) do neither intersect with the already examined projects of our preliminary qualitative study nor with the sample {\sf Stargazers}.\\
We followed the exact same approach to compute the data for the sample \textsf{SciC} as we did for the other sample. The results are listed in the second column in Table ~\ref{toolapplsamples:tab}.

\paragraph{Results for \textsf{SciC}}
A total of 4050 Java files and 548,976 lines of code were scanned in this sample. The tool detected 515 SP-formula code matches in eleven of 14 projects (78.5\%) in 199 different Java files and 1,794 lines of SP-formula code. Hence the density of detected SP-formula code based on files and lines of code is $\rho^{SP}_{files}=4.91\%$ and $\rho^{SP}_{LOC}=0.32\%$, respectively.
\attic{In Figure~\ref{fcmatchesinprojects:fig:scatter} we see that in this sample there are projects with very high numbers of matches, in particular three projects with 188, 151 and 140 matches, respectively.}
In Figure~\ref{fcdensityinprojects:fig} we can see that 11 of the 14 projects have a considerably higher SP-formula code density than
most of the projects in the \textsf{Stargazers} sample.
Furthermore, Figure~\ref{fcdensityinprojects:fig} reveals that six of the implemented patterns also occur in this sample. Note, that in contrast to the \textsf{Stargazers} sample, here the two patterns {\sf NFIAA} and {\sf NFIAS} 
have a high density.

\attic{

\paragraph{Discussion}
\attic{It is not surprising that the \textsf{TOP50P1} sample revealed many more SP-formula code matches compared to the samples \textsf{Random2} and \textsf{Stargazers}. Although it consists of far fewer projects\attic{ one-twentieth of the number of projects as \textsf{Random2} resp. one-half of that of \textsf{Stargazers}}, it is comprised of 1.35, resp. ten times as many lines of code. }

As can be seen in Figure~\ref{fcmatchesinprojects:fig}, the number of matches across the whole \textsf{Stargazers} sample can be described as follows:
\attic{variation of 
the number of matches over all projects is very similar for samples \textsf{Random2} and \textsf{Stargazers}:}
There is a small interquartile range, a median value of 2
and a small amount of outliers denoting many more matches compared to all other projects in the sample.\\
\attic{ In contrast, the interquartile range of {\sf Top50P1} is considerably
larger and its median much higher.}
\attic{Nevertheless, we found that the SP-formula code densities for all three samples are 
not very different.}
Next, we look at Figure~\ref{patterndens:fig}, which gives an overview of the number of occurrences (frequency) relative to lines of code of the SP-formula code patterns in Table~\ref{patterntypes:tab}.
For each sample, the bar chart shows the absolute number of matches for each of 
the SP-formula code patterns implemented by our tool
Note that the y-axis in Figure~\ref{} is scaled to the square-root of the frequencies. 
Furthermore, there are no bars for the patterns {\sf NFECC} and 
{\sf NFECS} in the chart, because no matches were found. We actually implemented these 
patterns as variations of the {\tt for}-loop patterns {\sf NFIAS} and {\sf NFIAA}, because we expected that programmers would use {\tt foreach} for similar purposes.
We see that three patterns, in particular \textsf{FIS} (844 \textsf{TOP50P1}, 641 \textsf{Random2}, 101 \textsf{Stargazers}), \textsf{FES} (749, 464, 50) and \textsf{FIA} (178, 115, 7) stand out in contrast to the others, especially over all investigated samples. In addition to these three patterns, the patterns \textsf{FEA} and \textsf{NFISS} also occur in all three samples. It is not surprising that every pattern occurs in sample \textsf{TOP50P1} but we see that all patterns also occur in sample \textsf{Random2} indicating that the derived patterns actually seem to be common in practice. Furthermore the patterns \textsf{NFESS}, \textsf{NFIAA} and \textsf{NFIAS} were not detected in the sample \textsf{Stargazers} which might be due to the smaller number of projects investigated in this sample. This and the overall small number of occurrences also leads to the impression that these patterns are actually rare and the discovery of those in the exploratory study (Section~\ref{keyword:sec}) is a result of the conducted specialized search for formula code in the first place. 
}

\attic{
	, no other pattern occurred in all of the three samples. Actually, there are no matches of the formula code patterns  in both samples, \textsf{Random2} and \textsf{Stargazers} as well as no occurrence of the pattern \textsf{FEA} in the sample \textsf{Random2}. This leads to the impression that these patterns are actually rare and the discovery of those in the exploratory study (Section~\ref{keyword:sec}) is a result of the conducted specialized search for formula code in the first place. }

\section{Discussion}
\label{discussion:sec}

\subsection{Diversity of Formula Code}
From our qualitative study, we can conclude that there exists a wide range of formula code. On that basis, we can give first answers to research question \textbf{RQ1}: One general observation is that the full extent of an implemented formula was often not directly recognizable. For some samples in our qualitative study, it took considerable effort to track down the complete implementation of a documented formula. Especially when the code fragments were split across different source code artifacts such as classes or files. We manually inspected 142 matches and classified 47 as real formula code. While the formula code was certainly diverse, almost half of the code involved \texttt{for}-loops, and also almost half of the code used arrays. We found several examples of incremental implementations of formulas. In these cases, the code may only implement the formula when we assume a certain dynamic behavior, such as a certain sequence of method calls or object instantiations. Detecting these kinds of formula implementations and asking the programmer to add assertions to the code to assure the correct dynamic behavior could help to prevent erroneous usage. Furthermore, reconstructing a formula in mathematical notation from the code was a very valuable and intuitive part in the process of code comprehension. From that point of view, it is obvious that one integral requirement for formula code support is the visual mathematical representation of the respective code. 
 
\attic{ with overall less symbols needed can be offered which is able to reduce mental load. }
\attic{SD: Beide Ansätze sind nicht ausreichend! Thus, static analysis will neither be sufficient to detect nor to infer those kinds of formula code. We conclude the necessity of dynamic analysis techniques to fully assess such types of formula implementations. }
\attic{From an outstanding position, we can only guess the developers intend to design such a formula code construct. We imagine the ability to add constraints to program code artifacts, e.g. methods and classes, defining necessary calling contexts in which a method or object should occur. 
Those constraints shell serve as invariants in order to assure a correct dynamic behavior and with that, a correct implementation of the intended formula. On the one hand this reminds of test frameworks, e.g. unit testing, and on the other hand of constraint programming. We see promising investigations in the symbiosis or integration of constraint and object oriented programming, albeit the paradigms differ. }

\paragraph{Splitting} We also found that very often, complex expressions are split into multiple partial computations storing the interim results in temporary variables --- either with
 names that intend to describe the computational part it represents, or with simple names such as \texttt{tmp}. The developers often followed an approach in which they split the expressions by operator precedence. For example, fractions the numerator and denominator computations are separated, stored in temporary variables and then divided in a subsequent statement. Besides the motivation of making the code more reusable and readable, developers seem to also split large expressions to facilitate debugging of the formula code. The splitting of computations makes it possible to leverage interactive break-point debuggers to investigate the interim results as well as assure correct application of arithmetic operations and functions according to their sequence and precedence. Current interactive debuggers only support line-by-line evaluation of code. A more fine grained approach in which single operations in the same line can be investigated seems helpful. For formula code, an interactive mathematical visualization where parts of a formula could be collapsed, expanded, evaluated, used as break-points and also be edited with respective effectual code changes would form a promising extension to current debugging mechanisms.
 
\attic{Furthermore, we think of an interactive mathematical visualization of the complete expression tree of those arithmetic computations. To offer this to a developer within the source code editor where components can be hidden, expanded, collapsed, evaluated and also be edited with respective effectual code changes would form a promising extension to current debugging mechanisms.}

\paragraph{Naming and Formatting}
The scope of arithmetic computations is not limited to scalar values. We often found groups of variables coherent in the way they were named and the kind of operations applied to them. These variables were actually used to represent a single element of a vector or matrix. We found these groups of variables being modeled either as object variables encapsulated in a class and thus programmatically specifying their coherence, or completely unbounded as local variables. 
\attic{The variable's naming either follows an indexing approach, e.g. \texttt{m\_00} for matrix value at row 0 and column 0, mostly for matrices or a direct naming of the components, e.g. \texttt{v\_x}, \texttt{v\_y}, \texttt{v\_z}, for vectors.}
The formatting of corresponding code snippets gives the impression that developers try to visually form a vector or matrix in a known mathematical way. For vectors, we often discovered a vertical arrangement such that every component of a vector is computed subsequently in its own line of code. For matrices, the variables are sub-grouped for every row and column in a similar approach as for vectors, i.e. for each row- and column vector separately. 
\attic{When arrays are used to model a matrix, we notice a two dimensional formatting of the variables representing the matrix's elements whereby rows and columns become visible.} 
This gives us even more evidence that developers want to have a visual representation
of the formula code close to its mathematical representation.

\attic{The numerical operations applied on the variables, i.e. in particular arithmetics and function calls, repeat for every element of such a group of variables. They only differ in the concrete variable or value used but the operations and their sequence were largely identical. This phenomenon represents a code clone of type 2. From a software engineering perspective it would make sense to refactor this into a method call. In the context of formula code, the developers explicitly didn't intend to apply these refactorings. We assume them intending to reduce computational overhead in form of method calls and unnecessary administrative code in order to increase the application's performance.\\}

\paragraph{Arrays} In total, 36 out of the 47 investigated code examples in our qualitative study were concerned with vector or matrix computations. 
Besides the low level modeling of vectors and matrices through semantic groups of variables, arrays are being used for this purpose.
\attic{The ability to combine multiple values under one variable and the opportunity to access those through an indexing mechanism, makes arrays a very appropriate way to model multidimensional mathematical structures like vectors and matrices. We discover both, arrays encapsulated in a class or without any further abstraction when used to store the elements of a vector or matrix.}
Surprisingly, also one-dimensional arrays are being used to model matrices. 
It is possible that developers want to avoid the computational overhead
involved with n-dimensional arrays. In Java an n-dimensional array
is actually a one dimensional array with pointers to (n-1)-dimensional arrays. Hence, we assume that developers intend to trade computing offsets for dereferencing pointers. Among others, this represents one phenomenon where we encountered a performance optimization at the expense of the formula's recognizability and thus overall code readability and maintainability.

\attic{Thus multiple array objects need to be instantiated and no consecutive memory region will be allocated. Accessing the elements can therefore happen to be more expensive compared to the calculation of the correct index (integer arithmetic) in a one dimensional array even though it might results in less readable code.}

\paragraph{Loops}
Along with the utilization of arrays comes the application of loops, not only to iterate through arrays, but also, and in particular, to implement sum and product formulas. Those kinds of formulas occurred not only in vector/matrix contexts but also in scalar computations and sequences.
In our small sample in Section~\ref{keyword:sec}, \texttt{for}-loops 
with an indexing variable were the predominant kind of loops used.
This is not surprising, since these are already syntactically 
very close to the $\sum$ or $\prod$ operators in mathematics.
Nested conditionals within the loops body could directly be translated to a part of the mathematical representation.
Only in a few cases the formula code contained extra code, for example debug statements.
\attic{In most cases of the discovered sum or product formulas there was no code involved but the code modeling the formula. It is rare but did appear that only a slice of an identified code snippet implemented the actual formula. 
The remaining code either had no interference with the formula or was debug code.}

\attic{Together with the loop's header, a precise predicate of which elements/values are traversed can be created and finds place in the mathematical sigma or pi notation. In those cases a mathematical representation already combines multiple code fragments to one predicate. This supports code comprehension tasks in reducing the time to understand what set is really being processed by a loop. 

If those visual representations were made interactive, we assume edits on a the mathematical notation, e.g. a predicate, resulting in an edit of the respective code as well. That way a developer can either work with the code or its mathematical representation to implement the formula. }

While the \texttt{while}-loop plays a secondary role in our findings of the qualitative study, the \texttt{foreach}-loop turns out to be relevant in the context of formula code as well. 
\attic{Every observation from the \texttt{for}-loop holds to the same extent with the one major difference:} \texttt{foreach}-loops are mostly applied on collections and in particular to traverse them where the sequence of processing the elements does not play a significant role. In the formula code context, \texttt{foreach}-loops can not only be used
to implement sum and product formulas over collections, but also to implement
logical formulas (predicates) related to these collections using the universal quantifier $\forall$ and existential qualifier $\exists$.
The derived code patterns for sum and product formulas of Section~\ref{patterns:sec} concentrate on \texttt{for} and \texttt{foreach}-loops in simple and nested variants.

\subsection{Frequency of Formula Code}
To answers research question \textbf{RQ2}, we decided to investigate the frequency of formula code for sums and products in terms of formula code densities as defined in Section~\ref{fcdensity:para}. These metrics are relative to the size of the projects and thus give a better impression than absolute numbers of matches and are comparable between projects.
Above all, we found a 7.4 times higher SP-formula code density in sample of scientific-computing projects compared to the one
of engineered software projects. Within the latter sample, we also found the SP-formula code-rich projects came from application domains 
related to scientific-computing.
 Nonetheless, the small size of the \textsf{SciC} sample calls for repeating the study
 as soon as more scientific-computing Java projects become available on GitHub.

\paragraph{Estimations}
Based on our detection tool's recall, we determined a rough estimation of SP-formula code density of $\widetilde{\rho}^{\,SP}_{\,LOC} = 0.14\%$ for the \textsf{Stargazers} sample and $\widetilde{\rho}^{\,SP}_{\,LOC} = 1.03\%$ for the \textsf{SciC} sample. In other words, we estimate
that about 1 of 700 lines of code in an engineered Java software project and 1 of 100 lines in a scientific-computing Java software project is part of an implementation of a sum or product formula.
\attic{ We take the results from the quantitative study as directive to start over with scientific-computing projects to discover more, and strengthen our current insights on the diversity of formula code. }
%
What does this mean in practice? 
The daily programming tasks of a software developer are not limited to writing code. Tasks related to the project's code base comprise writing, editing and, to a major proportion, reading and with that comprehending the code. Thus, we are certain that an average software developer gets in touch with SP-formula code multiple times a work week. 
\attic{Considering our qualitative study, which revealed also more diverse formula code fragments other than SP-formula code, the overall formula code densities are imaginable even higher. Software developers could therefore benefit greatly from tools supporting work with formula code. Thus, we expect that investigations on formula related source code have a promising future.\\}

\begin{figure}[ht]
	\centering
	\includegraphics[width=.9\textwidth]{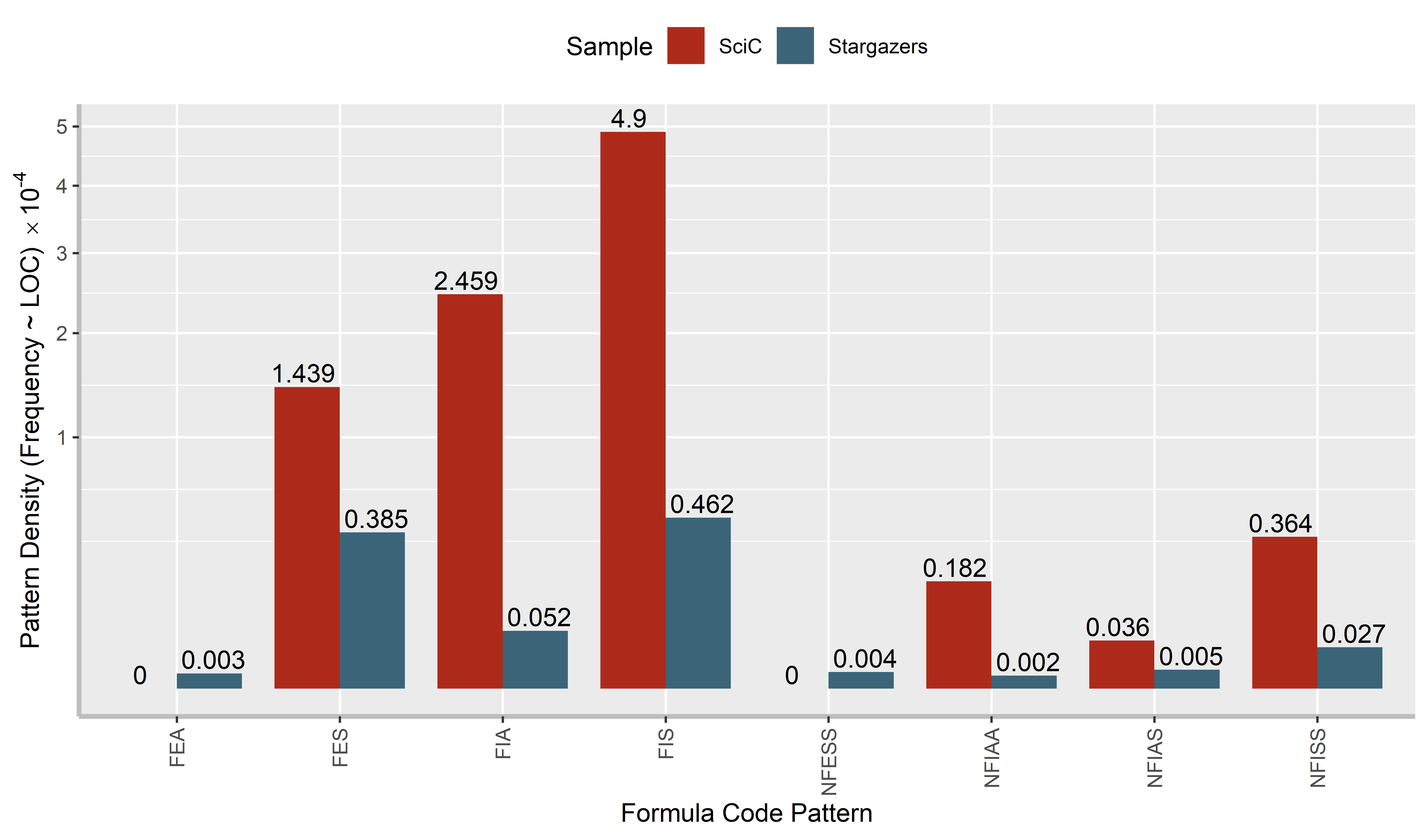}
 	\caption{\label{patterndens:fig}Density of SP-formula code matches for each pattern relative to LOC per sample.}	
\end{figure}

\paragraph{Patterns}
In our quantitative study, we have shown the relevance of all patterns in the context of formula code. Figure~\ref{patterndens:fig} reveals that in the \textsf{Stargazers} sample, every SP-formula code pattern was detected at least once. In the \textsf{SciC} sample, six of the eight patterns appeared.
Furthermore, the three SP-formula code patterns \textsf{FES}, \textsf{FIA} and \textsf{FIS} (all based on non-nested \texttt{for}-loops) are the most frequent patterns in both samples (Figure~\ref{patterndens:fig}). We determine a 3.7, 46.2 respectively 10.6 times higher density of these patterns in the \textsf{SciC} sample. Besides that, we discover higher densities for the nested SP-formula code patterns in the \textsf{SciC} sample as well. This represents another insight showing an increased relevance of the patterns as well as an increased probability to encounter SP-formula code in scientific-computing projects.
\attic{In addition to that, we suggest software development tools, e.g. visualizations for comprehension and debugging of formula code or documentation tools like the one from Moser et al.~\cite{Moser2015}, should address formula code utilizing simple non nested \texttt{for}-loops first. That way, tools would already cover a major part of SP-formula code and thus, would bring the most benefit at first.}

\paragraph{Percentage of loops implementing formulas}
To put the SP-formula code density into perspective, we investigated how many (nested) \texttt{for}-loops occur in the source code and
how many of these actually implement SP-formulas. To this end, we implemented two patterns using regular expressions, similar to the SP-formula code patterns, in order to detect simple (non-nested) and nested \texttt{for}- and \texttt{foreach}-loops. In the following, we use the term loops to refer to both \texttt{for}- and \texttt{foreach}-loops.
We applied our detection tool (Subsection~\ref{tool:sec}) with the new patterns to ensure comparability. Table~\ref{generalforloopsinsamples:tab} summarizes the detection tool's results for the general loop patterns. We report for both simple and nested loops the total number of matches in the whole sample (\#matches), the total number of files in which a match occurred (\#files) and the total number of projects in which a match occurred (\#projects).
The tool yielded 114.793 simple and 9,558 nested loops in the \textsf{Stargazers} sample. \attic{ in 38,143, respectively 5,278 files and in 790, respectively 412 projects of the \textsf{Stargazers} sample.}
Surprisingly, in 156, respectively 534 of the 946 projects, i.e. in 16,49\%, respectively 56.44\%, no simple, respectively nested loops were found. In contrast, 2,738 SP-formula code matches, based on simple loops and 120 SP-formula code matches based on nested loops were detected in this sample. Therefore, 2.38\% (every 42nd) of all simple loops and 1.25\% (every 80th) of all nested loops implement a sum of product formula according to our definition. When we apply a rough estimation depending on the detection tool's recall, 
in 7.71\% (every 13th) a simple and in 4.06\% (every 25th) a nested loop implements a SP-formula. 
\attic{With this insight, we assume it even more likely for a software developer to get in contact with SP-formula code.} 
In the \textsf{SciC} sample, we found 6,350 simple and 1,685 nested loops. \attic{ in 1,255, respectively 460 files and 13, respectively 10 of the total 14 projects.}
The SP-formula code matches in this sample amount to 483 simple and 32 nested loops. Thus, 7.60\% (every 13th), respectively 1.89\% (every 52nd) of all investigated simple, respectively nested loops form a SP-formula according to our definition. Taking the detection tool's recall into account, we roughly estimate that 24.60\% (every 4th) for simple, and 6.14\% (every 16th) for nested loops implement sum or product formulas. 

\attic{Thus, we conclude a higher frequency of SP-formula code and a greatly higher importance of the simple \texttt{for}-loops in the context of scientific-computing projects. Furthermore, the domain of scientific-computing represents the target domain and with it the target group of software developers to address with respective debugging and code comprehension tools. Apart from this, the code base and contributing developers of scientific-computing projects can be consulted as pool for real-world code examples and potential participants for conducting user studies. On the other hand requirements for debugging and code comprehension tools might differ among different developer specializations. To discover this, a user study could be designed so that the target group itself can be correlated with their requirements. Altogether, the domain of scientific-computing is very attractive for future investigations, in particular regarding the diversity and frequency of formula code in Java.}

\attic{Due to the small number of projects in the sample \textsf{SciC}, its considerably higher SP-formula-code density could be accidental. Figure~\ref{fcdensityinprojects:fig} reveals that all scientific-computing projects have a SP-formula code density of less than 0.01, which we considered to be the decision threshold to call a project SP-formula code-rich in the \textsf{Stargazers} sample.}

Overall, we think that it is reasonable to assume that formula code in scientific-computing is more frequent, more diverse and more complex than in most other application areas.\attic{ and that our findings support these hypotheses.}

\attic{Another subtle topic during this work is software performance. We encountered many code design decisions against readability and maintainability in favor of performance benefits. Also, we didn't detect any lambda expressions in the investigated code bases. Functional programming is in our conception often close to a mathematical notation. Therefore, we assumed to detect formula code examples utilizing lambda expressions. Thus, we imagine the absence of lambda expressions to implement mathematical formulas due to performance reasons. This suggestion forms yet another question to investigate in future work, either in the context of formula code or in general Java coding contexts.}

\begin{table}
	\caption{Results of the detection tool for each sample utilizing the general loop patterns.}
	\label{generalforloopsinsamples:tab}
	\centering
	\begin{tabular}{l|ccc|ccc|}
		\ & \multicolumn{3}{c|}{Simple \texttt{for}-loops}
		& \multicolumn{3}{c|}{Nested \texttt{for}-loops} \\
		\hline
		Sample
		&{ \sf \#matches }
		&{ \sf \#files }
		&{ \sf \#projects }
		&{ \sf \#matches }
		&{ \sf \#files }
		&{ \sf \#projects } \\
		\hline
		\sf Stargazers & 114,793 & 38,143 & 790 & 9,558 & 5,278 & 412 \\
		\hline
		\sf SciC & 6,350 & 1,255 & 13 & 1,685 & 460 & 10 \\
		\hline
	\end{tabular}
\end{table}

\todo{OM: conclusion resp summary of discussion in conclusion section? stuff we didn't find, e.g. lambda s etc...}

\subsection{Comments and Formula Code}
To investigate to what extent formula code is documented we manually inspected the 53 files of the oracle data set (Section~\ref{para:oracle}) that contained formula code fragments (see supplementary material~\cite{moseler_oliver_2018_1252324}). For nested formula code, we only recorded information about source code comments for the outermost level. We found, that 54\% of these 139 outermost formula code fragments were commented (8\% on the file or class level, 30\% on the method level and 16\% on the statement level), but only for 26\% of the formula code fragments the comments actually documented the semantics of the formula and we found that only 9\% of the comments provided non obvious information that was helpful for understanding the formula code. In most cases comments that related to the semantics simply described the result like ``{\tt // get longitude}'' at the statement level or both the result and the parameters in the method header using JavaDoc. In few cases, comments documented not only what is computed, but also how it is computed. Only three comments referred to external resources for documentation: one to a book, one to Wikipedia and one by mentioning the name of the formula. Many of the other comments were related to development tasks, e.g. ``{\tt //FIXME: This needs a much more efficient implementation}''. 

Distinguishing the types of formula code, we found that
20\% of formula code that did not involve loops (simple arithmetic)  had comments that were helpful for understanding the code. At only 6\% resp. 5\%, this ratio was much lower for simple and double nested loops. Thus, at least in our oracle data set, SP-formula code was rarely commented in a way that helped to better comprehend the implemented formula.

\begin{table}
	\caption{Properties of the comments in 139 formula code fragments 
	         of the oracle data set~\label{codedcomments:tab}}
	\centering \small
	\begin{tabular}{l|c|c|c|c|}
	    & & \multicolumn{3}{c|}{\bf Type of formula code}\\
	   
        & {\bf all} 
          &{\bf non-nested loop}
          &{\bf nested loop}
          &{\bf simple arithmetic}\\
		\hline
	     {\bf Formula code fragments} & 139 & 82 & 21 & 35 \\
	    \hline
	    {\bf Comments in these fragments} & 75 & 41 & 12  & 21 \\
	   \hline
	  {\bf Comments that ...} & \multicolumn{4}{c|}{}\\  
	  \hline
      {\bf document the semantics} & 36 & 17 & 4 & 15 \\
      {\bf are helpful for program understanding} & 13 & 5  & 1 & 7 \\
      {\bf refer to an external source} & 3 & 0 & 1 & 2\\
	\end{tabular}
\end{table}

\todo{Auswertung helpful comments}
\todo{Tabelle mit Ergebnisse zu Comments}

\subsection{The Need for Tool Support}
The scarcity of helpful source code comments is a first indication that there is 
a need
for tools to help developers better understand formula code. To answer this question from a programmers' point of view, we conducted a small online survey with computer science students.
The survey consisted of four bug finding tasks in real-world formula code and a socio-demographic questionnaire. We recorded accuracy as well as completion time for each task.

\begin{figure}[htb]
	\centering
	\caption{Task 4 of the online survey.\label{task4:fig}}
	\fbox{
	\includegraphics[width=.82\textwidth]{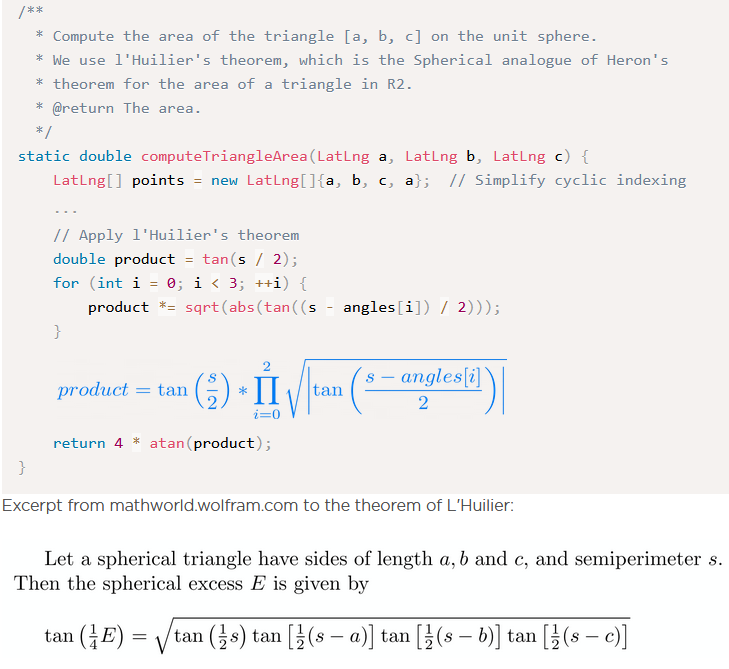}
	}
\end{figure}

For each task, we presented a task page with the source code of the defective formula implementation 
and additional information (as described below) to the participants (see Figure~\ref{task4:fig}). To finish the task, they had to answer with yes or no if they were able to find the error in the code.
We used the time span between showing the task page with the source code and answering with yes or no as the task completion time. On the next page the participants had to describe the error. With the help of that description we were able to determine whether they really found the error in the formula code. 
Finally, we asked all participants whether they thought, that a tool for reconstructing mathematical formulas from program code would be helpful? They had to answer this question with yes or no, and could provide additional remarks, if they wanted to.

For the tasks we selected four formula code fragments from GitHub that we already
investigated in our qualitative study (Section~\ref{keyword:sec}). 
We chose two simple code fragments (Set~A: trace of a matrix (Task 1) and scalar product (Task 2)) 
and two more complex fragments (Set~B: Chi-Square test of independence (Task 3) 
 and l'Huilier's theorem (Task 4)).
For tasks in Set B, additional documentation  from Wikipedia resp. MathWorld 
about the implemented formula was shown on the task page. We manually injected defects
into the formula code in a way such that the tasks in the same set were of approximately
the same difficulty. Furthermore, each  
participant had to solve one task in each set with resp. without the reconstructed formula displayed
close to the formula code on the task page. We used a within-subject design as shown in Table~\ref{studydesign:tab} with the reconstructed formula as the factor of the experiment. We provide a print version of the entire online survey for both groups and the corresponding results in the supplementary material~\cite{moseler_oliver_2018_1252324}.

\begin{table}
	\caption{Study design: task pages with and without reconstructed formula~\label{studydesign:tab}}
	\centering \small
	\begin{tabular}{l|c|c|c|c|}
        & \multicolumn{2}{c|}{\bf Set A} & \multicolumn{2}{c|}{\bf Set B} \\
        & Task 1 & Task 2 & Task 3 & Task 4 \\
        \hline
{\bf Group 1} &   w    & w/o    & w      &  w/o \\
{\bf Group 2} & w/o    & w      &  w/o   &   w\\
   \end{tabular}
\end{table}


%


We sent invitation mails to 266 computer science students. 
To motivate study participation, we raffled three \euro{\,20} vouchers among the participants.
In total, 27 of these actually completed the survey, after data cleaning based on a minimum total task completion time of 60 seconds 17 participants (9 bachelor, 6 master, and 2 PhD students) remained in the data set (with a mean task completion
time of 236 seconds and a mean duration of completing the survey of 21 minutes).

\todo{demographic data}


\begin{table}
	\caption{Mean accuracy scores for all participants, participants with 
	  resp. without affinity for mathematics~\label{scoresAB:tab}}
	\centering \small
	\begin{tabular}{l|c|c|}
	   \multicolumn{3}{c}{All}\\
	  \cline{2-3}
	   & With & Without \\
	  \hline
	  Set A & 0.53 & 0.76 \\
	  Set B & 0.24 & 0.35 
    \end{tabular}
\hfill
   	\begin{tabular}{l|c|c|}
	   \multicolumn{3}{c}{Math}\\
	   \cline{2-3}
	   & With & Without \\
	  \hline
	  Set A & 0.56 & 0.78 \\
	  Set B & 0.11 & 0.22 
    \end{tabular}
\hfill
    \begin{tabular}{l|c|c|}
	   \multicolumn{3}{c}{No Math}\\
	   \cline{2-3}
	   & With & Without \\
	  \hline
	  Set A & 0.50 & 0.75 \\
	  Set B & 0.38 & 0.50 
    \end{tabular}
\end{table}

For each participant we computed the sum of correctly solved tasks as a accuracy score for each set of tasks. The mean scores are shown in Table~\ref{scoresAB:tab}. As expected, the scores for Set~B were 
lower than those for Set~A, actually about half the size, indicating that the tasks 
in Set~B were more difficult. Unfortunately, when comparing the two treatments, with and without reconstructed
formula, we found that for both sets the participants performed better without the reconstructed
formula. This finding did certainly not meet with our expectations and maybe due to
the small number of participants. This assumption is backed by the even more
unexpected finding that for more difficult tasks (Set~B) students with affinity to mathematics (9 of 17 in this data set)
performed much worse than the rest. Here we used the socio-demographic data provided
by the students whether they chose mathematics as an advanced course at high school 
or selected mathematics as the minor subject at university as an indicator 
of their affinity to mathematics.
Also, if we only look at Task 2, 85.5\% 
of the students in Group 2, but only 77,7\% in Group 1 solved this task correctly.
In other words, for Task 2 the treatment with the reconstructed formula 
had higher accuracy. 


In total, 16 out of 17 participants found it helpful to have a tool that reconstructs mathematical  formulas from the source code. Six of  these provided an additional remark. We coded their remarks and identified three reasons why they found such a tool to be helpful:
\begin{description}
\item [\it Math rich software projects:] One participant stated that such a tool would be helpful in ``(...) software projects such as simulation software where many (physical) formulas have to be implemented''. 
\item [\it Testing formula code:] Four participants suggested that a tool that reconstructs a mathematical formula from the source code is helpful on assessing the correctness of the implementation. 
\item [\it Defect detection in formula code:] One participant found it ``frustrating to not find a defect in formula code which really can become very complex'' and that ``such a tool certainly can simplify that''.
\end{description}

\attic{
 [4] Ja, aber wie hilfreich es ist hÃ¤ngt sicherlich von der Anwendung ab. Entwickelt man z.B. Simulationssoftware, so mÃ¼ssen viele (z.B. physikalische) Formeln implementiert werden. Auch fÃ¼r stark theoretische bzw. mathematische Anwendungsgebiete wird es hilfreich sein. Arbeitet man hingegen eher im Bereich Softwaretechnik und -Design, wo man sich mehr mit der allgemeinen Struktur eines groÃŸen Softwareprojektes beschÃ¤ftigt, sind die konkreten Implementierungen einer bestimmten Formel nicht so relevant.                                                                                                                                                                                                                                     
 [6] Es ist sehr frustrierend die Fehler nicht zu finden, was bei mathematischen Formeln echt komplex werden kann und mit Sicherheit vereinfacht werden kann.                                                                                                                                                                                                                                                                                                       
[10] Es passiert Ã¶fters, dass man versucht eine mathematische Formel als Programmcode umzuschreiben und die verschiedenen Schritte falsch miteinander verknÃ¼pft, was das Endergebnis dann natÃ¼rlich falsch werden lÃ¤sst. Ein Werkzeug welches einem dabei hilft mathematische Formel korrekt zu Codieren wÃ¤re eine Bereicherung.                                                                                                                                                                                                                                                                                                                                      
[13] Ein Werkzeug zur Rekonstruktion mathematischer Formeln wÃ¼rde helfen, fremden Programmcode leichter zu verstehen und kÃ¶nnte dem Programmierer bei der Implementierung mathematischer Formeln zeigen, ob seine Implementierung korrekt ist.                                                                                                                                                                                                                                                                                                                    
[14] Ich bin mir nicht genau sicher, was mit so einem Werkzeug gemeint wÃ¤re und fÃ¼r welche Einsatztwecke es gedacht ist. Es klingt aber zuminest fÃ¼r PrÃ¼fzwecke nÃ¼tzlich.                                                                                                                                                                                                                                                                                                                                                                                   
[16] Ich kann nicht sagen, ob solch ein Werkzeug hilfreich ist, da dies in dieser Evaluierung nicht genutzt habe und mir nicht vorstellen kann, wie es aussieht, oder sich wÃ¤hrend des Programmierens verÃ¤ndert. Der Task den ich hier evaluiert sehe, ist Fehlerfinden in mathematischem Programmcode ohne UnterstÃ¼tzung einer IDE und wahrscheinlich noch die Java-Skills der Probanden (denn wer kennt schon so etwas wie LatLng).                                                                                                                                                                                                                                                                                                                               
[17] Nachdem ich die Beispiele durchgegangen bin, halte ich ein solches Tool fÃ¼r sehr sinnvoll.  Man kÃ¶nnte diese Frage auch Leuten stellen, die nicht an der Studie teilgenommen haben, um festzustellen, wie das mit den Formeln wahrgenommen wird.    Ein solches Tool ist sicher super.  Aber man sollte auch die umgekehrte Richtung in Betracht ziehen.  D.h. ich habe bereits eine mathematische Formel implementiert und mÃ¶chte nun wissen, ob diese valide ist.  DafÃ¼r kann man sich solch einen Code erstellen lassen und ihn dann mit dem Geschriebenen vergleichen.  Das ist fÃ¼r Typ-2-Klone (Gleichheit bis auf Variablennamen) ohne Probleme mÃ¶glich. DafÃ¼r kann man sich die Arbeit von Brenda S. Baker ansehen (Parameterized Pattern Matching).

}

\todo{conclusion}

Our results show that with increasing complexity of the code (Set~B vs Set~A), the performance considerably decreased. Thus, a tool should focus on the 
more complex cases to address this performance gap.
While the performance results in our survey did not provide sufficient and clear 
evidence that 
showing the reconstructed formula next to the formula code is sufficient
to support understanding of formula code, the vast majority of
the participants agreed that a tool for reconstructing formulas would be helpful.


\section{Limitations}
\label{limitations:sec}

To classify a code fragment as formula code, we do not require that the programmer's original intention was to implement a mathematical formula, but instead it is sufficient that the implemented computation could be expressed by a mathematical formula. On the other hand, some program code fragments that really implement a mathematical formula might not directly be expressible in a mathematical notation since the respective code is already too far away from the maths, possibly due to performance optimizations, refactorings or applied coding hacks.

Each of the empirical studies presented in this paper comes with its own limitations. 

\paragraph{Qualitative Analysis of Formula Code Examples}
The formula code examples that we manually inspected were
selected based on keywords occurring in their comments.
Undocumented formula code, i.e. code that had none of
our keywords in its comments, may have different properties.
Furthermore, while our selection of keywords certainly introduces
a bias toward sum and product formulas, it did 
not affect the oracle data set, since we annotated any 
formula code that we found. We also tried to
increase credibility and transferability by following
established coding methods and by discussing all reconstructed
formulas in group sessions with all authors. This also
holds for our coding of the application domains.


\paragraph{Quantitative Tool Evaluation}
While the tags that we used to annotate the oracle data set were based on the results of the qualitative study, we did not exclude any kind of formula code only because it was not considered in the qualitative study. However, the annotation of the oracle data set depends on the subjective assessment whether some code fragment implements a formula.

\todo{OM: the github topic is freely assignable. there is no confidence that a given topic really states that a project is concerned with it.}

\paragraph{Quantitative Analysis of SP-formula code on GitHub}
We tried to carefully distinguish between the SP-formula code detected by
our tool and the SP-formula code actually present in a sample. We  
used a randomized, stargazer-based sampling strategy to increase the generalizability
of our findings to engineered software projects on GitHub.  
The generalizability of our results for the sample \textsf{SciC} 
is strongly limited by the small size of the sample.

\attic{ In particular, this holds for the density of detected formula code;  our estimation of
the general formula code density is an informed guess, based on
the assumption that the distribution of those kinds of formula code,
that are not detected by our tool, is similar in GitHub to the one
in the oracle data set.
}



To enable other researchers to verify our results, we provide all data that is not protected by copyright (e.g. content of GitHub repos) as supplementary material~\cite{moseler_oliver_2018_1252324}.


\section{Conclusion}
\label{conclusion:sec}
So far, research in software engineering has focused on the synthesis but not the analysis of formula code. In this paper, we first presented a qualitative study designed for gaining insights into the diversity of formula code in real world Java projects (\textbf{RQ1}). We found that code that developers document as formula code has special properties. The observed phenomena range from coded mathematical notation in the names of variables, complex arithmetic operations split by a debugging or precedence strategy, groups of variables with coherent naming and coherently applied operations, \attic{and thus forming vector or matrix computations,} over sum and product formulas based on \texttt{for}-loops up to incremental formula implementations that depend on the dynamic behavior (call sequence). 
In particular, we find it promising to provide an alternative representation of the respective code in terms of a mathematical notation. During our qualitative study, deriving such a mathematical representation of the code supported the process of code comprehension to a great extent. If those code visualizations are made interactive and reachable directly from within the source code editor, we assume 
that they would greatly facilitate debugging of formula code~\cite{vissoft:2020}. Designing and implementing such debugging features and assessing their usability and efficiency are part of our plans for future research.

Furthermore, we presented an approach to detect SP-formula code using syntactic patterns in combination with a set of constraints on the variables occurring in the matched code fragments. We derived these patterns based on our preliminary qualitative study and evaluated the effectiveness of our approach in terms of recall and precision. On that basis, we performed a case study to investigate the frequency of SP-formula code in a sample of 1,000 open source Java stargazer projects on GitHub (\textbf{RQ2}). We also looked at the application domains of  the SP-formula code-rich projects and found a major overlap with scientific, respectively technical subject areas, e.g. information retrieval, signal processing, computer networks, statistics, machine learning and simulation. This inspired us to also apply our tool to a sample consisting solely of scientific-computing projects (\textsf{SciC}). Since they give a more realistic impression on the real distribution of SP-formula code in open source Java projects, here, we only give densities estimated based on our detection approach's recall. The absolute numbers have been presented and discussed in the previous sections. We estimate that one of 700 lines of code in the \textsf{Stargazers} and even one of 100 lines of code in the \textsf{SciC} sample is part of an implementation of a sum or product formula. In addition to the line-based metrics, we also computed the total number of simple and nested \texttt{for}-loops and \texttt{foreach}-loops in the samples. We estimate that in the stargazers sample every 13th simple respectively 25th nested loop implements a sum or product. In the \textsf{SciC} sample the ratio is considerably higher: every 4th simple respectively 16th nested loop.
Based on these numbers, it is reasonable to assume that an average software developer will have to write, or at least comprehend, SP-formula code multiple times a work week.  

With two small studies, we explored the potential of specialized tools for formula code.
First, we investigated the source-code comments of 139 real-world formula code fragments. We found that the comments rarely supported the code comprehension task of the implementation of a formula. This lack of helpful comments could be mitigated by tools for (semi-)automated source-code commenting or documentation of formula code are promising research objectives. 
Second, in an online survey, we examined whether showing reconstructed mathematical formulas next to the formula code supports the defect detection task. To this end, we selected four real-world formula code fragments of different complexity. The performance results in our survey did not provide clear evidence that showing the reconstructed formula next to the formula code supports defect detection, and with that the code comprehension task on formula code. Except for one, all participants agreed that a tool for reconstructing formulas from source-code would be helpful. They found tools for formula code helpful especially applied on defect detection and verification resp. testing tasks while implementing a mathematical formula as well as in maths rich software projects.

\attic{Thus, specialized tools should definitively address that kind of code in order to cover common and frequent examples at first. We were not able to test our approach against other existing tools, in particular the RbG tool by Moser et al.~\cite{Moser2016,Moser2015}. They applied their tool on industrial closed source software mostly written in Fortran and C/C++. However, from the examples provided online\footnote{\textsf{http://codeanalytics.scch.at/}}, which consist mostly of arithmetic operations and \texttt{for}-loops, we assume their patterns to cover similar formulas.}

\attic{Future effort for us will be the detection tool's extension and enhancement in terms of patterns it can detect, e.g. vector and matrix patterns, and increased recall. Since our approach is a static detection tool and we discovered also formula code examples which depend on dynamic program behavior, we also think of a tool which is able of a dynamic formula code classification. We imagine such a tool to particularly being useful in debugging and thus code comprehension contexts.\\}

As we determine a 7.4 times higher SP-formula code density in the \textsf{SciC} sample, we think that the design of tools and language features specialized for formula code in
 this domain is a promising route for future research.
 Thus, we intend to enhance and extend the patterns in our tool, e.g., by adding vector and matrix patterns, and in particular, we want to investigate other programming languages such as Python, which are more common in the field of scientific-computing. 
 
\attic{Besides that, the domain of scientific-computing offers the opportunity to address experts in the field of implementations of mathematical formulas, and thus forms a pool for potential participants of user studies. Since the topic of software performance plays an omnipresent role concerning formula code and due to the discovered opportunities to facilitate the debugging of that special kind of code, we plan on investigating bug reports in order to explore possible correlations between formula code and bug- density or types, e.g. performance bugs.}

\todo{OM: Future Work, user studies to investigate the usefulness(?) of the mathematical representation in the program code.}

\todo{OM: Future Work, debugging features, inline debugging, interactive visualizations, expression trees...}

\todo{OM: participants for user studies from oss scientific-computing projects}

\todo{OM: correlation between formula code and (performance)bugs?}

\todo{OM: Java features like lambda expressions didn't occur but are actually considered closer to math notation than regular for loops, ref?}

\todo{OM: Review2 question 4: are code patterns derived from java applicable to python?}

\todo{consider experts in the field, MATLAB... projects}

\todo{Diversity and frequency of
formula code in Python}

\todo{Bugs and formula code, e.g. fix-inducing changes}
 
\todo{tool support for debugging}

\bibliographystyle{elsarticle-num-names}
\bibliography{formula.bib}







\end{document}